\documentclass[preprint2]{emulateapj}

\usepackage{gensymb}

\usepackage{subfigure}
\newcommand{\comment}[1]{}

\slugcomment{accepted to The Astronomical Journal}
\shortauthors{Kunder et~al.}
\shorttitle{Kinematic analysis of BH~261}

\begin{document}

\title{The Milky Way Bulge extra-tidal star survey: BH~261 (AL~3)}

\author{
Andrea Kunder\altaffilmark{1},
Zdenek Prudil\altaffilmark{2},
Kevin Covey\altaffilmark{3},
Joanne Hughes\altaffilmark{4},
Meridith Joyce\altaffilmark{5,6},
Iulia T. Simion\altaffilmark{7},
Rebekah Kuss\altaffilmark{1,8},
Carlos Campos\altaffilmark{1},
Christian I. Johnson\altaffilmark{9},
Catherine A. Pilachowski\altaffilmark{10},
Kristen A. Larson\altaffilmark{3},
Andreas J. Koch-Hansen\altaffilmark{11},
Tommaso Marchetti\altaffilmark{2},
Michael R. Rich\altaffilmark{12},
Evan Butler\altaffilmark{1,13},
William I. Clarkson\altaffilmark{14},
Michael J. Rivet\altaffilmark{1},
Kathryn Devine\altaffilmark{15},
A.~Katherina~Vivas\altaffilmark{16},
Gabriel I. Perren\altaffilmark{17},
Mario Soto\altaffilmark{18},
Erika Silva\altaffilmark{3}
}

\altaffiltext{1}{Saint Martin's University, 5000 Abbey Way SE, Lacey, WA, 98503, USA}
\altaffiltext{2}{European Southern Observatory, Karl-Schwarzschild-Strasse 2, 85748 Garching bei M\"{u}nchen, Germany}
\altaffiltext{3}{Department of Physics \& Astronomy, Western Washington University, MS-9164, 516 High St., Bellingham, WA, 98225}
\altaffiltext{4}{Physics Department Seattle University, 901 12th Ave., Seattle, WA 98122, USA}
\altaffiltext{5}{Konkoly Observatory, MTA CSFK, Budapest, Konkoly Thege Mikl\'os \'ut 15-17, Hungary}
\altaffiltext{6}{MTA CSFK Lend\"ulet Near-Field Cosmology Research Group, 1121, Budapest, Konkoly Thege Mikl\'os \'ut 15-17, Hungary}
\altaffiltext{7}{Shanghai Key Lab for Astrophysics, Shanghai Normal University, 100 Guilin Road, Shanghai, 200234}
\altaffiltext{8}{Department of Mathematics, Oregon State University, 1500 SW Jefferson Way, Corvallis, OR 97331}
\altaffiltext{9}{Space Telescope Science Institute, 3700 San Martin Drive, Baltimore, MD 21218, USA}
\altaffiltext{10}{Indiana University Department of Astronomy, SW319, 727 E 3rd Street, Bloomington, IN 47405 USA,}
\altaffiltext{11}{Zentrum f\"ur Astronomie der Universit\"{a}t Heidelberg, Astronomisches Rechen-Institut, M\"{o}nchhofstr. 12-14, 69120 Heidelberg, Germany,}
\altaffiltext{12}{Department of Physics and Astronomy, UCLA, 430 Portola Plaza, Box 951547, Los Angeles, CA 90095-1547, USA}
\altaffiltext{13}{Department of Astronomy, University of Washington, Physics-Astronomy Bldg, Room C319, Box 351580, Seattle, WA 98195-1700}
\altaffiltext{14}{Department of Natural Sciences, University of Michigan-Dearborn, 4901 Evergreen Rd. Dearborn, MI 48128, USA,}
\altaffiltext{15}{The College of Idaho, 2112 Cleveland Blvd Caldwell, ID, 83605, USA}
\altaffiltext{16}{Cerro Tololo Inter-American Observatory/NSF’s NOIRLab, Casilla 603, La Serena, Chile}
\altaffiltext{17}{Instituto de Astrof\'isica de La Plata, IALP (CONICET-UNLP), 1900 La Plata, Argentina}
\altaffiltext{18}{Instituto de Astronom\'ia y Ciencias Planetarias, Universidad de Atacama, Copayapu 485, Copiap\'{o}, Chile}

\begin{abstract}
The Milky Way Bulge extra-tidal star survey (MWBest) is 
a spectroscopic survey with the goal of identifying stripped globular cluster stars from inner Galaxy 
clusters.  In this way, an indication of the fraction of metal-poor bulge stars that originated from globular 
clusters can be determined.  We observed and analyzed stars in and around 
BH~261, an understudied globular cluster in the bulge.
From seven giants within the tidal radius of the cluster, we measured 
an average heliocentric radial velocity of $<$RV$>$ = $-$61$\pm$2.6~km~s$^{-1}$ 
with a radial velocity dispersion of $\rm <\sigma>$ = 6.1$\pm$1.9~km~s$^{-1}$.  
The large velocity dispersion may have arisen from tidal heating in the cluster's orbit about 
the Galactic center, or because BH~261 has a high dynamical mass as well as a 
high mass-to-light ratio.  From spectra of five giants, we measure an average metallicity of 
$\rm <[Fe/H]>$ = $-$1.1$\pm$0.2~dex.  We also spectroscopically confirm an RR Lyrae 
star in BH 261, which yields a distance to the cluster of 7.1$\pm$0.4~kpc.  
Stars with 3D velocities and metallicities consistent with BH 261 reaching to $\sim$0.5 degrees from 
the cluster are identified.  A handful of these stars are also consistent with 
the spatial distribution of that potential debris from models focussing on the most recent disruption of the cluster. 
\end{abstract}

%\keywords{Stellar populations(1622) --- Galactic archaeology(2178) --- Milky Way dynamics (1051) --- Galactic bulge(2041) --- Galaxy bulges(578) --- Globular star clusters(656)}

\section{Introduction} 
\label{sec:intro}
The connection between Globular Clusters (GCs) in the inner Galaxy and the hierarchical growth 
of the Milky Way is still largely unknown.  One reason bulge globular clusters (GCs) are difficult to 
place into proper context within our Galaxy's formation is that they often have features not seen in the 
GC halo or disk population.  For example, Terzan~5 and Liller~1 are bulge GC fossil fragments that host an 
old ($\sim$12~Gyr) and young ($\sim$1-3~Gyr) stellar population \citep[e.g.,][]{ferraro21}.  
NGC~6441 and NGC~6388 are bulge GCs with abnormal horizontal branches -- too blue and 
extended for their $\rm [Fe/H]$ metallicities and with abnormal frequencies and pulsation 
properties in their RR Lyrae populations \citep{pritzl00, pritzl01}.  
%NGC~6522 is postulated to 
%have neutron-capture elements consistent with harboring an early generation of fast 
%rotating stars \citep[][but see 
%also Ness et al.~2014]{chiappini11, barbuy21a}.  
The Galactic bulge is home to the most metal-rich 
GCs in our Galaxy, and studies of the elemental abundances (e.g., O and Na) in 
bulge GC stars indicate that the evolution of many bulge GCs are not similar to 
that of the halo \citep[e.g.,][]{munoz17}.  

Especially the metal-poor stars in the field of the bulge appear to be connected to inner Galaxy GCs.    
Field stars with $\rm [N/Fe]$ over-abundances are thought to be former members of a population 
of GCs that was previously dissolved and/or evaporated \citep[e.g.,][]{schiavon17, fernandeztrincado21}.
This is the same mechanism that has been shown to donate stars to the halo \citep[e.g.,][]{martell11, koch19}.  
Stripped GC stars are also contenders for the origin of the double RC feature in the bulge, as 
it has been shown that the chemical abundances of the stars in the X-shaped bulge are consistent with having been formed from GC fossil remnants \citep{lim21}. 
Although it is expected that Galactic GCs lose mass through processes like evaporation and tidal stripping 
\citep[e.g.,][]{leon00, baumgardt03, moreno14, baumgardt21}, the extent of stripped GC stars in the bulge 
is unclear.  Yet in the bulge, where dynamical friction is much higher than in the halo, this process is likely 
a significant mechanism of the make-up of the bulge field, especially for the metal-poor bulge population.  

One hindrance in being able to draw connections between inner Galaxy GCs and place this population  
into context with Milky Way formation, is that these are dense systems in a crowded, extinguished 
part of the Milky Way, so observational analysis of inner Galaxy GCs is difficult.  Many of the inner 
Galaxy GCs are understudied, with basic parameters such as radial velocities and metallicities 
being undetermined, prompting new spectroscopic survey's to target inner Galaxy GCs \citep[e.g.,][]{saviane12, dias16, kunder21, geisler21}. 
The Milky Way Bulge extra-tidal star survey, MWBest, has the goal of spectroscopically  
identifying stripped globular cluster stars from inner Galaxy clusters.  
Globular cluster stars can and will escape close to the tidal boundary of the cluster as it moves 
through the inner Galaxy, influenced by the tidal force of the Milky 
Way, but the detection of stripped globular cluster stars in the inner Galaxy is limited in 
number \citep[e.g.,][]{gnedin97, meylan00, kunder14, kunder18, minniti18, kundu19}.  
The tidal force inflates the cluster (tidal heating), and tidal stripping removes mass in 
its outer region.  Due to the severe crowding of the bulge,
we concentrate on potential extra-tidal stars that lie a few tidal-radii 
($\sim 1-5 {\times}~r_t$) away from the cluster center.  

This paper focuses on the bulge globular cluster BH~261.  
\citet{andrews67} list it as AL~3, 
\citet{vandenbergh75} list it as BH~261, and \citet{lauberts82} list it as ESO456-SC78.  
The first color-magnitude diagrams (CMDs) of this 
region by \citet{carraro05} show very little evidence that it is a true cluster.  
It was the photometry presented in \citet{ortolani06} 
that allowed this cluster to be confirmed as a GC, and provided a photometric distance, reddening and 
metallicity from isochrone fitting.  
A deep CMD of BH~261 is presented by \citet{cohen18}, 
who confirm a sparse, blue horizontal branch morphology using the $Hubble~Space~Telescope$ (HST).  
Since then, new photometry from {\it Gaia} \citep{gaiacollab16, gaiacollab21} and VVV \citep{minniti10}
has been presented by \citet{gran22}, finding a photometric distance that places the cluster 50\% further away, 
on the far-side of the bulge, and finding a photometric metallicity that is a factor of 10 more metal-poor. 

Spectroscopic studies of stars in BH~261 include those by \citet{baumgardt19}, \citet{barbuy21} 
and \citet{geisler23}, who report an average radial velocity of $-$29.4~km~s$^{-1}$, $-$57.9$\pm$4.3~km~s$^{-1}$, 
and $-$44.9$\pm$3.8~km~s$^{-1}$, respectively.  One reason for these differing results may be the 
small sample sizes ($\sim$3 stars in each study), or it could be that BH~261 has a larger 
velocity dispersion then is able to be reported with the small sample sizes.  
The spectroscopic $\rm [Fe/H]$ of BH~261 is measured to be between $\sim$$-$1.0 and 
$\sim$$-$1.3 \citep{barbuy21, geisler23}, also based on three member stars.

The spectroscopic observations presented here allow a more detailed dynamical study to be carried out, since our observations extend out to $\sim$2$^\circ$ from the cluster center, or $\sim$20 tidal-radii.  
In \S2 the new data collected is described, and the radial velocities and metallicities are presented in \S3.  
In \ref{sec:feh} the extra-tidal stars identified are compared to theoretical predictions of tidal debris from the initial conditions of BH~261, and a comparison between BH~261 and other bulge GCs is carried out.   The conclusions are in \S4.

\section{Observations and Data Reductions} \label{sec:data}
\subsection{Target selection}

\begin{figure}
\centering
\mbox{\subfigure{\includegraphics[width=8.2cm]{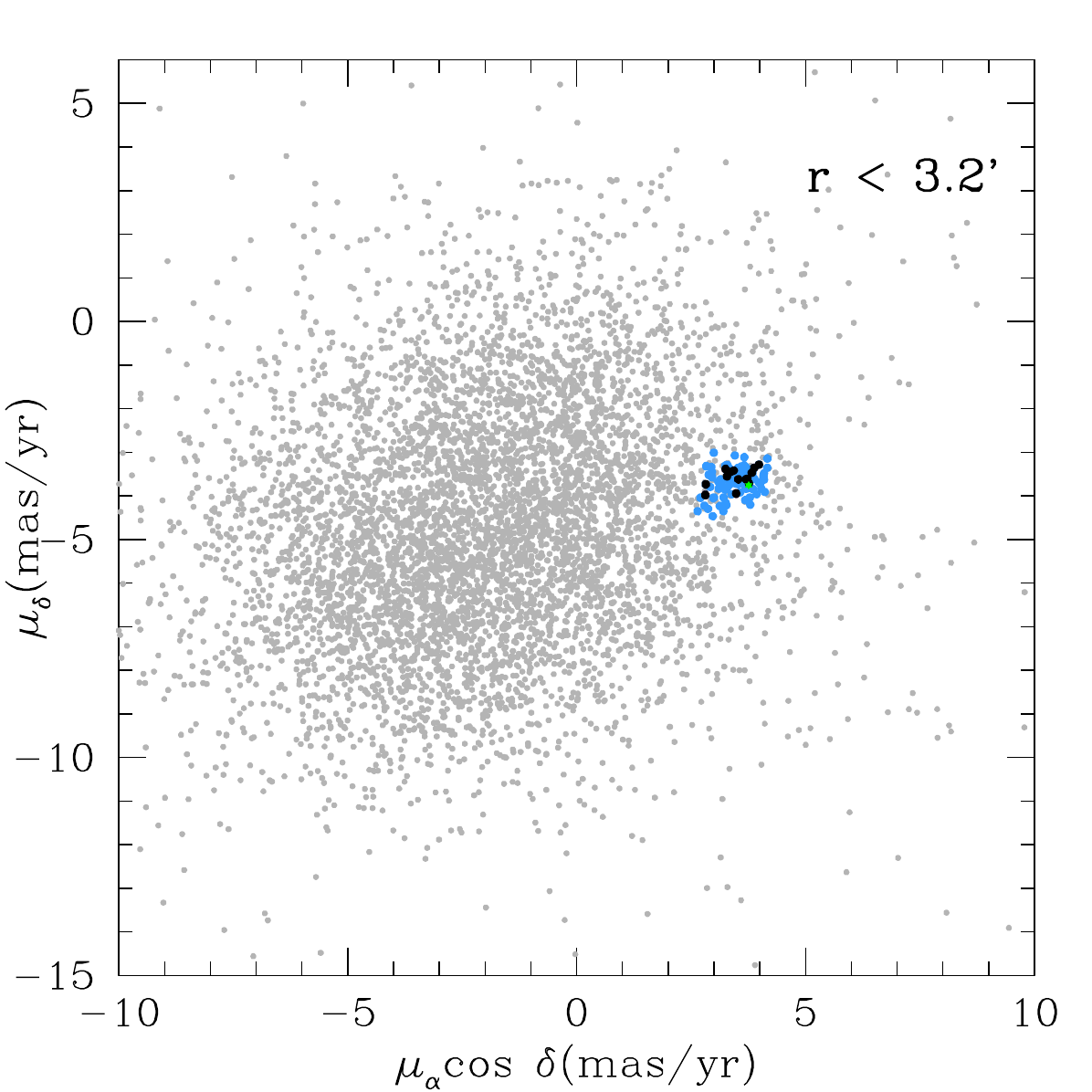}}}
{\subfigure{\includegraphics[height=8.2cm]{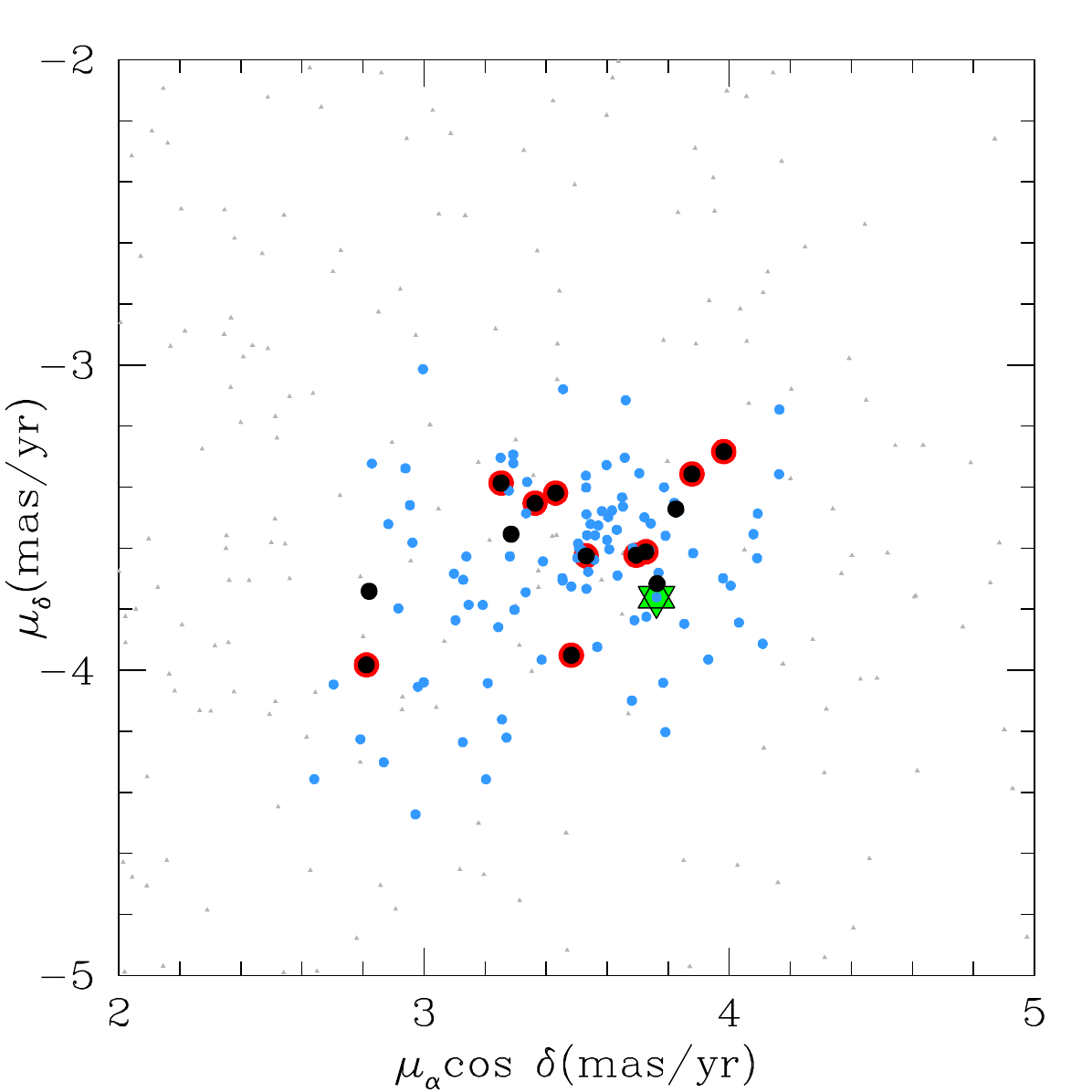}}}
\caption{
The {\it Gaia} proper motion distribution of the BDBS stars centered in a 3.2 arc minute 
radius from BH~261.  Blue points show stars with proper motions 
within 1.5 mas~yr$^{-1}$ in both  $\mu_\alpha$ 
and $\mu_\delta$ of the mean proper motion 
of the cluster, as well as those with stars with parallax values smaller than 0.4~mas.  
The black points indicate stars that were spectroscopically targeted, and those
with red highlights are those that were found to be radial velocity members.  
The star symbol (green) indicates the RR Lyrae star OGLE-BLG-RRLYR-35078.
}
\label{pms}
\end{figure}

BH~261 is heavily contaminated by both Galactic disk stars as well as bulge field stars, which 
makes efficient target selection difficult.  
The $Gaia$ catalog \citep{gaia22} was used to apply proper motion and parallax criteria to select stars consistent with the cluster and to cull both foreground and field stars, as shown in Figure~\ref{pms}.  
In particular, stars with proper motions within $\mu_\alpha \pm$1.5~mas~yr$^{-1}$ and $\mu_\delta \pm$1.5~mas~yr$^{-1}$ of the mean proper motion of the cluster were selected, where the mean proper motion of BH~261 is $\mu_\alpha$cos$\delta$ = 3.589$\pm$0.022~mas~yr$^{-1}$, $\mu_\delta$=$-$3.570$ \pm$0.020~mas~yr$^{-1}$ \citep{vasiliev21}.  
Stars with parallax values larger than 0.4~mas were discarded, as it was shown that these stars in general are part of the foreground disk \citep{marchetti22}.  
Proper motion and parallax information is not precise enough for cluster membership of BH~261; our derived radial velocities are ultimately used to select the most likely cluster members.

The Blanco DECam Bulge Survey (BDBS) catalog \citep{rich20, johnson20} was also used to 
select targets for the cluster, selecting stars with $u$ and $i$ photometry that 
would, in principle, encompass the cluster’s red giant branch (RGB) and blue horizontal branch (BHB).  
BDBS is a photometric survey covering more than 200 square degrees of the Southern Galactic 
bulge using the $ugrizY$ filters on the Dark Energy Camera.  
Photometry of approximately 250 million unique sources is available in BDBS, spanning 
the Galactic longitude range from l=$-$10$^\circ$ to +10$^\circ$ and the Galactic latitude range 
from b=$-$3$^\circ$ to $-$10$^\circ$. 
% {\bf A BDBS stack of the $g$, $r$, and $z$ filters show the cluster BH~261 in Figure~\ref{clarkson}.

%\begin{figure}
%\centering
%\mbox{\subfigure{\includegraphics[width=8.2cm]{bh261_4x2.png}}}
%\caption{
%A stacked BDBS image in  the $g$, $r$, and $z$ filters covering a 4' x 2' region centered on BH~261.
%}
%\label{clarkson}
%\end{figure}

\begin{figure}
\centering
\mbox{\subfigure{\includegraphics[width=9.2cm]{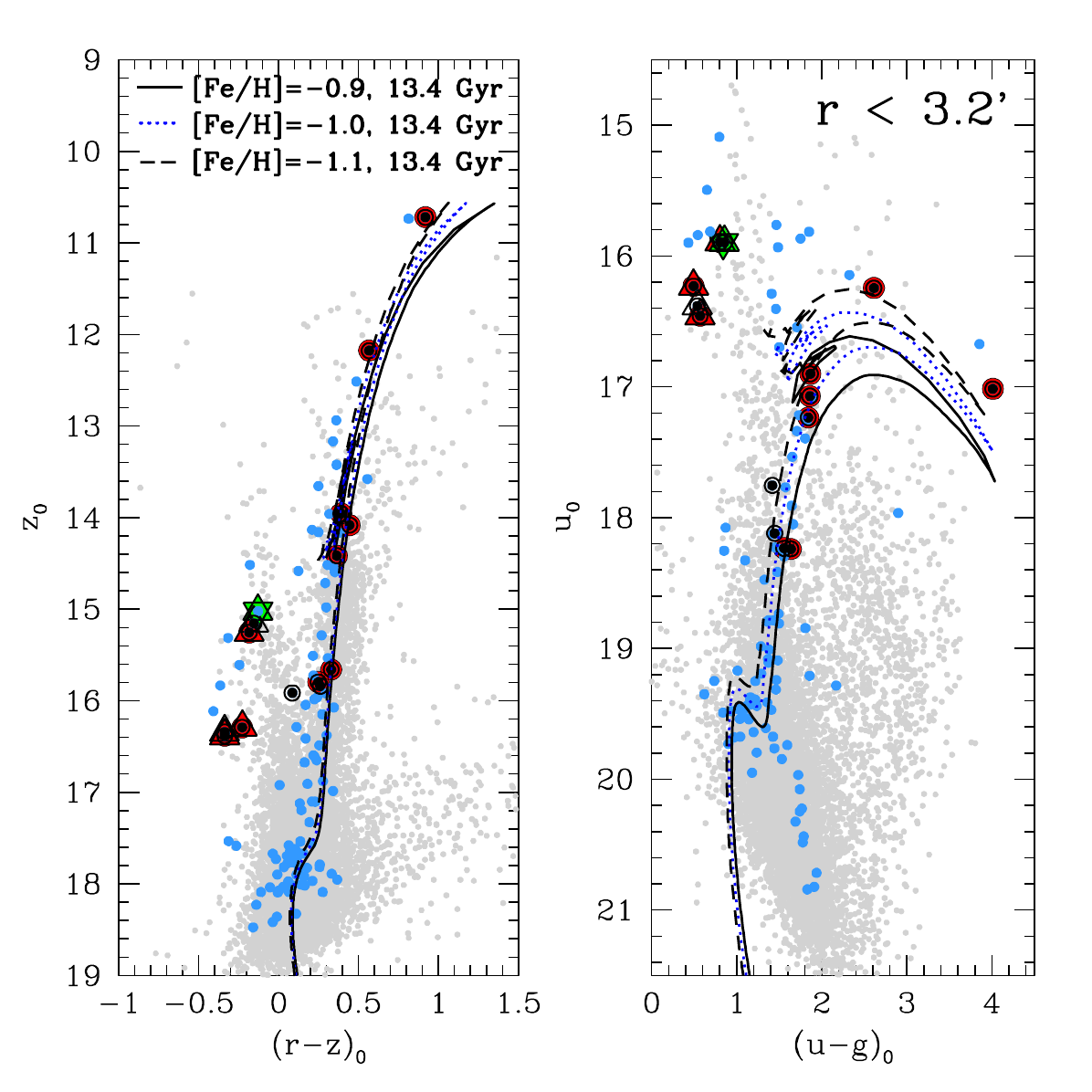}}}
\caption{
 The BDBS color-magnitude diagram showing the horizontal branch stars (triangles), 
giants (circles) and RR Lyrae star (star) for which radial velocities have been determined 
from AAOmega@AAT.  The filled red symbols indicate those stars that have radial 
velocities consistent with BH~261 (see Table~2), whereas the open triangle and circles 
are those stars that do not have velocities consistent with the cluster.  The underlying 
BDBS stellar distribution in this field is shown with small grey points and those with proper motions 
consistent with the cluster are shown in blue.  
The black lines show the MIST \citep[MESA Isochrones and Stellar Tracks,][]{choi16}
isochrone from which the cluster's distance is derived in this work 
(7.1~kpc). 
}
\label{cmd}
\end{figure}

The stars assigned the highest priority were those within 
the tidal radius of BH~261 that were consistent with being blue horizontal branch (BHB) stars.  
Because the BHB is more offset from the bulge field population (see Figure~\ref{cmd}), 
this should maximize the number of bonafide BH~261 stars.  However, BHB stars have a 
hotter temperatures than giants and red giants, so the proximity to and systematic blueward 
offset of the calcium infrared 
triplet to the hydrogen Paschen lines complicates stellar parameter determination.

We also targeted red clump stars 
with photometric metallicities more metal-poor than $\rm [Fe/H] = -$0.3~dex.  
\citet{johnson20, johnson22} show that  \hbox{\it u--i\/} colors can be used to obtain 
color-$\rm [Fe/H]$ relations for red clump stars good to $\sim$0.2 dex.  This precision 
is comparable to that of most spectroscopic metallicities of bulge stars \citep[see also][]{lim21}.  
The targeted stars have 
BDBS $u$-band photometry with formal uncertainties of $u_{err} <$0.024~mag, with the 
observed stars with $u \sim$18~mag or brighter having $u_{err} <$0.01~mag.  

\subsection{Observations and Reduction}
\begin{figure}
\centering
\mbox{\subfigure{\includegraphics[width=8.2cm]{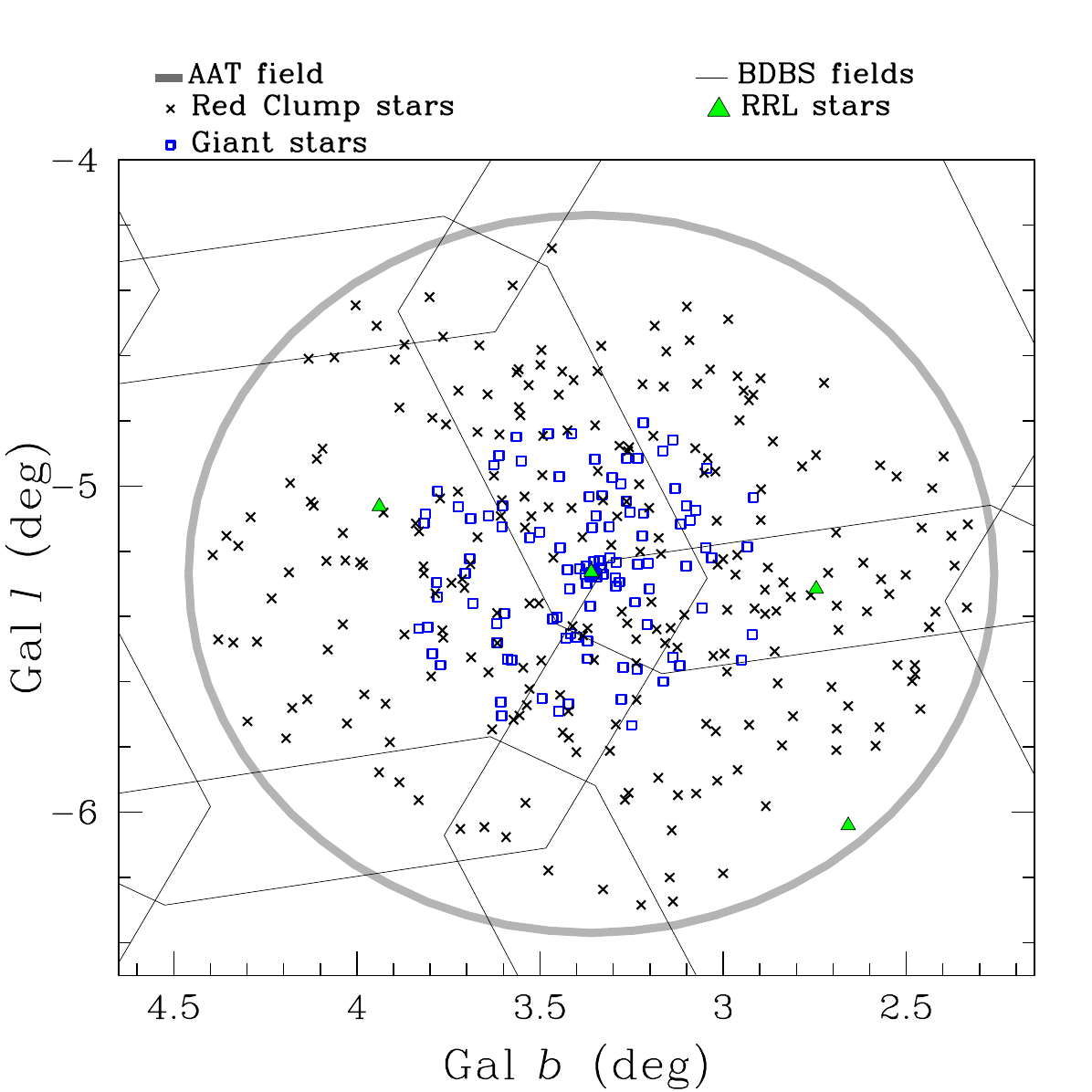}}}
{\subfigure{\includegraphics[width=8.2cm]{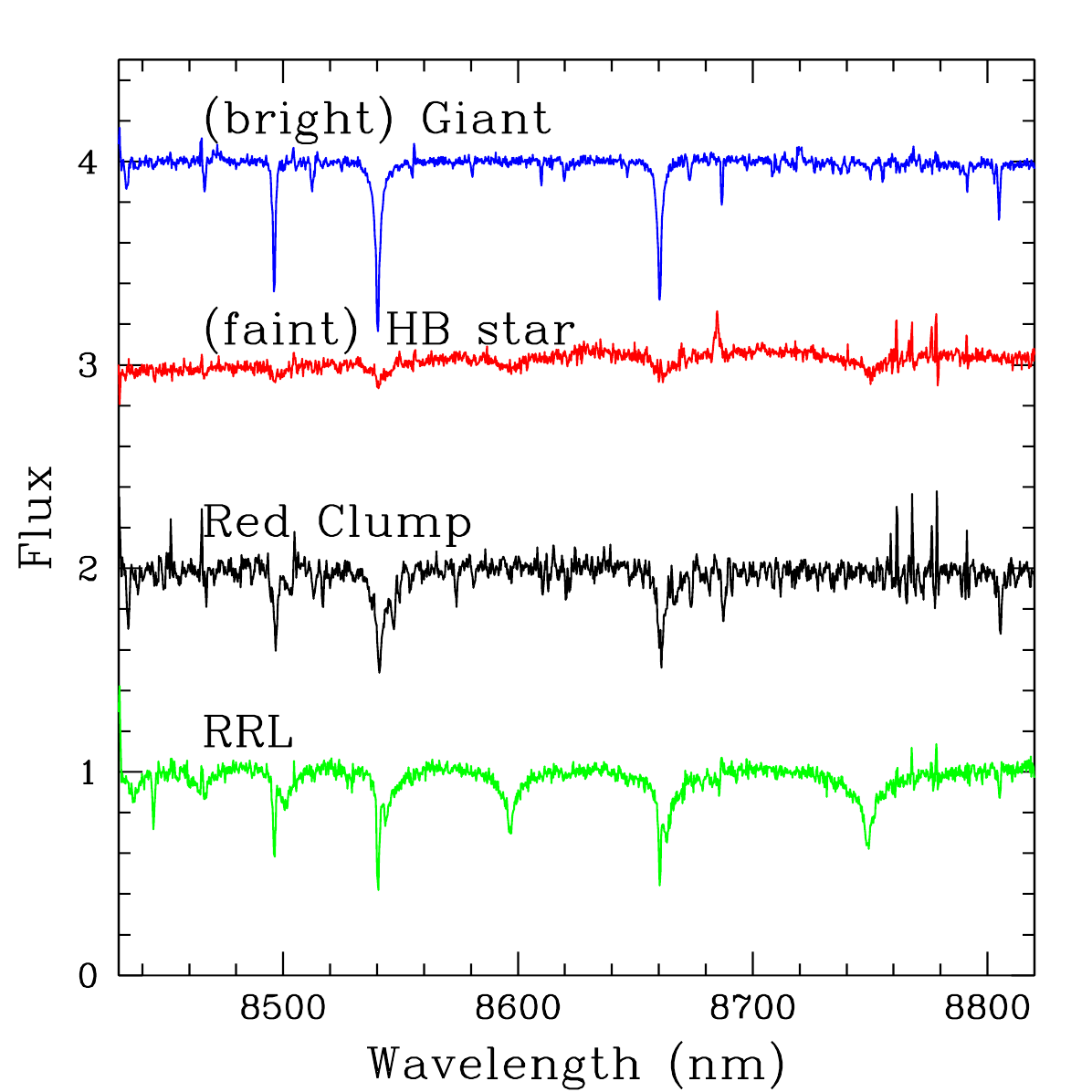}}}
\caption{
{\it Left:} The stars in and around BH~261 targeted spectroscopically with AAOmega@AAT.  The field is centered on 
the GC BH~261. 
{\it Right:} Example spectra from AAOmega illustrating the difference in quality between a bright giant ($i$=14.678~mag), a faint HB star ($i$=16.814~mag), the hot RRc star ($i$=15.538~mag) and a typical red clump star ($i$=16.515~mag).  
The spectra have been normalized and are offset for clarity. 
}
\label{spatial}
\end{figure}

New spectra were collected using the AAOmega multifibre spectrograph at the 3.9 m 
Anglo-Australian Telescope (Siding Spring Observatory, Coonabarabran, NSW, Australia).  The 
five night run occurred 20 July - 24 July, 2022 (PROP-ID: O/2022A/3002).  Plate configurations for the Two 
Degree Field (2dF) fibre positioner contained a combination of RR Lyrae stars, red clump stars and giants 
centered on the cluster and filling the 2 degree field of view, as shown in Figure~\ref{spatial}.  All of the 
stars targeted have proper motions consistent with BH~261.  The field was observed twice with two 
different configurations -- different giant stars were observed in both configurations to maximize number of 
potential cluster stars and extra-tidal stars, but the same red clump stars were observed as they are fainter, 
incase the spectra needed to be stacked.  Also, the same RR Lyrae stars were observed in each 
configuration to maximize phase coverage for these pulsating variables.  

A dual setup was used to employ the red 1700D grating, centered at 8600 \AA~and the blue 2500V grating, 
centered at 5000 \AA. In this manner, the easily seen calcium triplet (CaT) lines in the red were observed, and for 
the brightest stars, the Mg line at 5180 \AA~in the blue was prominent.  
This paper uses only the red part of the spectra, and any analyses of metallicities and $\rm [Mg/Fe]$ will be 
presented at a later stage.  The exposure times ranged from 4x30 min to 2x30 min, adjusting for weather and 
observing conditions.  The typical signal-to-noise was $\sim$5 per pixel for the fainter horizontal branch stars 
and $\sim$45 per pixel for the brighter giants.

The bias subtraction, cosmic ray cleaning, quartz-flatfielding, wavelength calibration via arc-lamp exposures, 
sky subtraction using dedicated sky fibers, and optimal extraction of the science spectra were carried 
out using AAO’s 2dfdr pipeline \citep{aao15}.  The final wavelength range is 8350--8800 \AA, with slight variations 
depending on the exact position of the spectra on the CCD. 

\subsection{Radial velocities and $\rm [Fe/H]$ measurements}
Radial velocities were measured using IRAF's {\tt xcsao} routine \citep{tody86, tody93} which utilizes cross-correlation against 
another spectrum.  The three spectra we used as cross-correlation templates were stars observed during 
the same run, selected from the Apache Point Observatory Galaxy Evolution 
Experiment \citep[APOGEE,][]{eisenstein11} database.  
In particular, APOGEE~2M18134674-2926056 (RV=27.88$\pm$0.03), 
APOGEE~2M17514997-2906055  (RV=$-$187.33$\pm$0.02) and APOGEE~2M17521244-2919510
(RV=65.13$\pm$0.05) were adopted as radial velocity templates.  This led to a median velocity error of 
$\sim$3~km~s$^{-1}$ for the giants and 9~km~s$^{-1}$ for the fainter and hotter HB stars.

\begin{figure}
\centering
\mbox{\subfigure{\includegraphics[width=8.2cm]{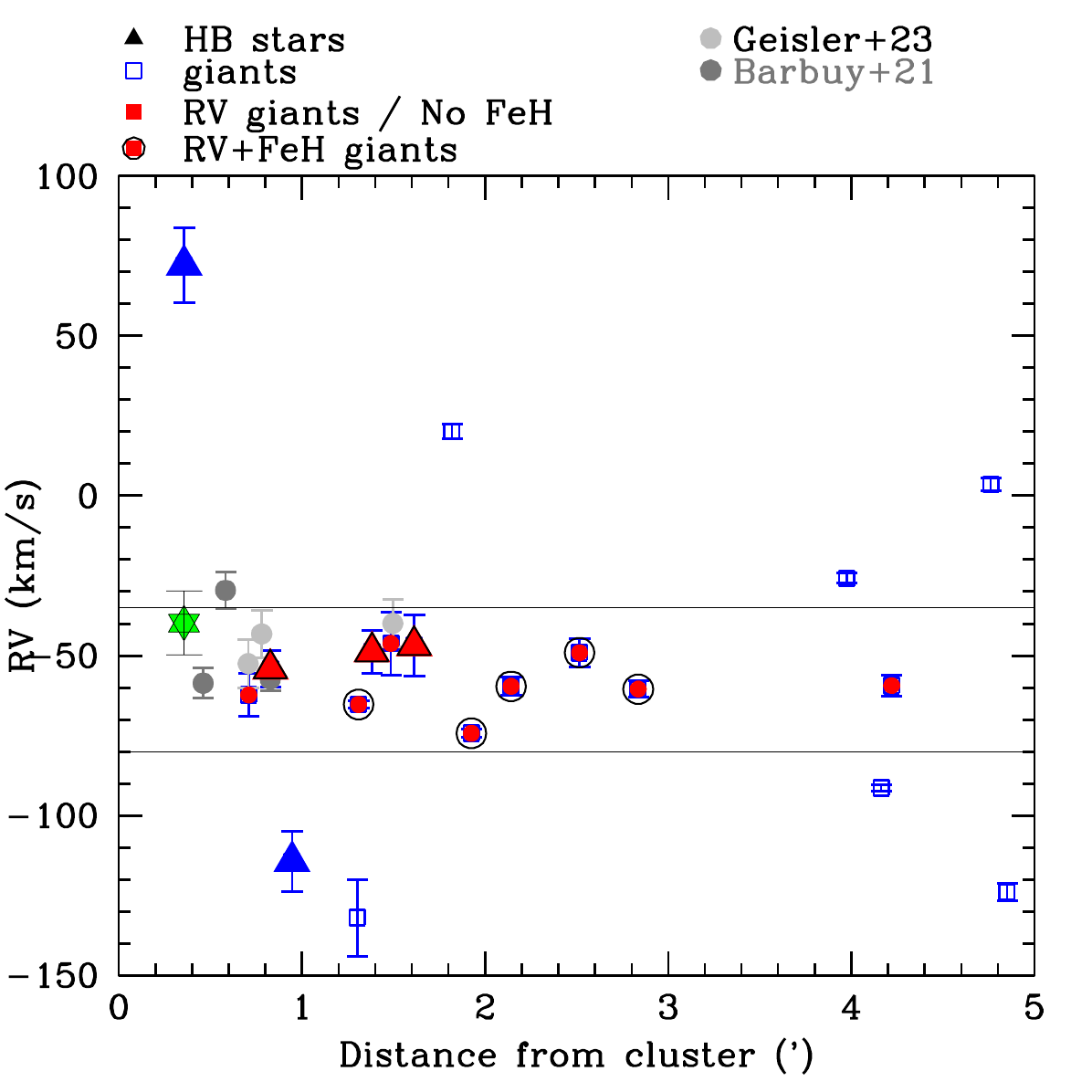}}}
{\subfigure{\includegraphics[height=6.5cm]{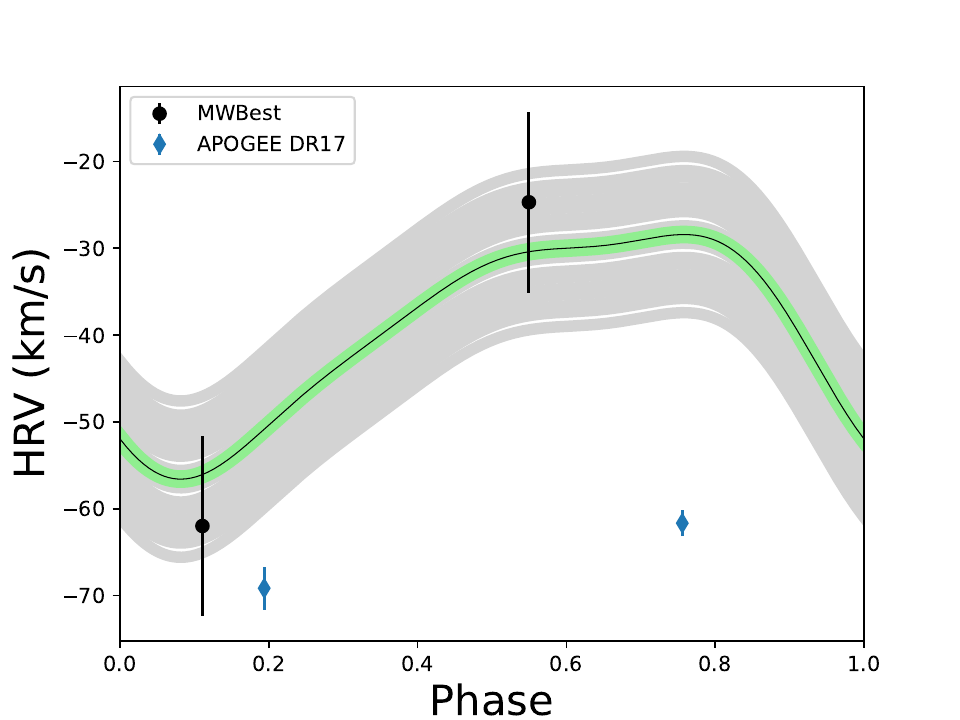}}}
\caption{
{\it Left:} The heliocentric velocities of our targeted stars within 5 arc-minutes of 
BH~261.  
The clump of twelve stars with radial velocities of $\sim$$-$60~km~s$^{-1}$ within 
the 3.5 arcmin of the cluster are consistent with being BH~261 stars.
{\it Right:}  The radial velocity curve of the first-overtone RR Lyrae, OGLE-BLG-RRLYR-35078, which has 
both a proper motion and a radial velocity consistent with BH~261.  The RRc template from \citet{prudil23} 
is used to obtain a center-of-mass radial velocity.  The APOGEE DR17 observations are also 
shown, but not used in the radial velocity determination.
}
\label{GC_RV_zoom}
\end{figure}

Two epochs of observations were collected for the RR Lyrae star OGLE-BLG-RRLYR-35078, 
and these are shown in Figure~\ref{GC_RV_zoom} (right panel).  Each epoch of observation is 
phased using the OGLE pulsation ephemerides and pulsation period.  To calculate the mean radial 
velocity, the RRc template presented in \citet{prudil23} is adopted.  The photometric scaling relation 
adopted between the photometric amplitude, $\rm Amp_V$, 
and line-of-sight velocity amplitude, $\rm Amp_{los}$, is $\rm Amp_{los}$ = 54(1) $\times$ $\rm Amp_{V}$. 
This was derived specifically for RRc pulsators from five well-sampled local RR Lyrae stars 
observed by APOGEE \citep{prudil23}. 
For $\rm Amp_{V}$, the OGLE $I$-Amplitude is transformed to $\rm Amp_{V}$ using 
$\rm Amp_{V} =$1.72 $\times$ $\rm Amp_{I}$ \citep{kunder13, prudil23}.  
The determined systemic velocity and its uncertainty is 
$-$39.8$\pm$12.4 km~s$^{-1}$.  
The 12.4 km~s$^{-1}$ uncertainty in the systemic velocity comes from adding in quadrature the 9.8~km~s$^{-1}$ individual radial velocity uncertainty to the 7.6~km~s$^{-1}$ uncertainty from the model fitted to find the systemic velocity.  %(7.6**2 + 9.8**2)
The source of uncertainty comes from the faint magnitude of this star combined with its hotter temperature.   
The shaded grey in Figure~\ref{GC_RV_zoom} about the scaled RRc template designates a 9.8~km~s$^{-1}$ uncertainty.  

The APOGEE DR17 catalog contains a number of RR Lyrae star observations, 
including two epochs of observations of OGLE-BLG-RRLYR-35078.  The 
{\tt allVisit-r12-l33.fits} file was used to 
extract the exact time each observation was taken as listed in the 
column {\tt JD}.  The Julian Date from this file refers to the middle 
of an exposure sequence, which is determined from exposure-time weighted 
mean of the mid-exposure times.  Again using the OGLE time of maximum brightness and 
OGLE period, the phase of each APOGEE RRL observation was calculated.  
The APOGEE radial velocity observations give a systemic velocity of 
$-$67.2$\pm$2.3~km~s$^{-1}$.  
The APOGEE mean velocity as well as that derived here are both consistent with the range 
of velocities seen for stars in BH~261.  Therefore, the radial velocity also confirms 
OGLE-BLG-RRLYR-35078 is a cluster member, in agreement with its proper motion and spatial 
proximity to the cluster.  

The radial velocity of OGLE-BLG-RRLYR-35078 presented here, $-$39.8$\pm$12.4 km~s$^{-1}$, 
is used throughout the paper, as the Calcium Triplet lines are stronger 
than the spectral lines in the APOGEE $H$-band wavelength regime, 
especially at the hotter temperatures of first-overtone RR Lyrae stars.  
Further, APOGEE's reduction pipeline \citep{nidever15} stacks all spectra of a given object to increase 
SNR and a radial velocity on the stacked spectra is determined. This radial velocity is a first 
estimate, or a base for prior, for further radial velocity determination for that given star.  
This procedure may be sub-optimal for stars that change their radial velocities with amplitudes 
of $\sim$15-50~km~s$^{-1}$ in short periods, like RR Lyrae stars.  The small formal 
uncertainties in the APOGEE spectra of 2.2~kms$^{-1}$ and 1.5~kms$^{-1}$ are 
almost certainly underestimated given the signal-to-noise of 4.1 and 7.6, respectively.  
As far as we know, this is the first publication using the APOGEE DR17 measurements of 
RR Lyrae stars, and we look forward to further discussion of APOGEE radial velocities for 
RR Lyrae stars in potential forthcoming papers. 

The SP\textunderscore ACE code \citep{boeche16, boeche21} was utilized for the determination of 
$\rm [Fe/H]$ metallicities for the giants.  For the red clump stars observed, photometric metallicities were 
calculated from the calibration between DECam passbands and $\rm [Fe/H]$ as presented in 
\citet{johnson20, johnson22}.  Spectroscopic metallicities from SP\textunderscore ACE are used 
to verify the authenticity of the red clump $\rm [Fe/H]$ metallicities.  The spectroscopic 
metallicities are discussed in more detail in \ref{sec:feh}.

\section{BH~261 Results} \label{sec:bh261}
\subsection{Distance}
The distance to BH~261 has been determined from CMD fitting and ranges from 
$d_\odot$ = 6.0$\pm$0.6 kpc from optical CMD fitting \citep{ortolani06, barbuy21} to 
$d_\odot$ = 9.12~kpc from infrared CMD fitting \citep{gran22}.  There is unfortunately no DR3 
parallax or kinematic distance for BH~261 \citep{baumgardt21}.  The reason for the large 
discrepancy in distance from CMD fitting is that there is a degeneracy between metallicity 
and distance:  adopting a $\rm [Fe/H] \sim -$1.0 
leads to a distance of $\sim$$d_\odot$ = 6.0~kpc whereas adopting a more metal poor  
 $\rm [Fe/H] \sim -$2.5 leads to a distance of $d_\odot \sim$ 9.5~kpc.  
 %Without a spectroscopic metallicity, knowing which $\rm [Fe/H]$ value to use is difficult.  
 %{\bf Even with a published spectroscopic metallicity of $\rm [Fe/H] \sim -$1.0, \citet{gran22} favor the larger distance and argue 
% 
The advantage of adopting a more metal-poor value for BH~261 is that a larger distance to the 
cluster makes it easier to explain its large velocity dispersion despite its low luminosity, as this 
gives a mass-to-light ratio more in line with what is seen for typical GCs.  
The recent spectroscopic $\rm [Fe/H]$ metallicities for BH~261 in \citet{geisler23} 
agree with the metallicity put forward by \citet{barbuy21} and \citet{ortolani06}, making it unlikely that the 
further distance is appropriate.
 
We report a RR Lyrae star --  OGLE-BLG-RRLYR-35078 -- that both lies 0.4' from the center of 
the cluster, and has a proper motion consistent with BH~261.  Our derived radial velocity further 
indicates it is a cluster member.  As such, this star can provide an independent indicator to estimate 
the distance to BH~261.  

Empirical period absolute magnitude metallicity (PMZ) 
relations for RR Lyrae stars have greatly improved, especially since the trigonometric 
parallaxes measured by {\it Gaia} have been released.  For the determination of the distance to this 
RRc star, the newly calibrated PMZ relations in \citet{prudil23} are employed.  
Briefly, the \citet{prudil23} calibrating dataset consists of 100 RR Lyrae stars with mean intensity 
magnitudes, reddenings, pulsation properties, iron abundances, and parallaxes from {\it Gaia} DR3.  
Both RRab and RRc pulsators are included in the calibrating set and it was shown that their 
derived PMZ relations accurately estimate the distance moduli to NGC~6121, NGC~5139, 
the LMC and SMC, as well as the the prototype of RR Lyrae class, RR Lyr.
Because the motivation behind the \citet{prudil23} PMZ relations is to use them 
for RR Lyrae stars toward the Galactic bulge, special care is given to calibrate the relations 
to the OGLE and VVV photometric system directly.  Further, a homogeneous metallicity scale 
is used for the calibrating sample that allows the direct use of photometric metallicity derived 
from the OGLE $I$-band photometry.  The \citet{prudil23} PMZ relations are used to investigate 
the distance to BH~261.  
Assuming $A_k$ = 0.04 $\pm$ 0.02 mag, and a photometric metallicity of 
$\rm [Fe/H]=-$1.25 \citep{dekany21}, a distance of d=7132 $\pm$312~pc is derived from 
the OGLE $I$-band and the VVV $K_s$ band.  Using only the VVV $J$ and $K_s$ bands, 
$A_k$ = 0.08 $\pm$ 0.09 mag is found, and a distance of d=7006 $\pm$427~pc is derived.  
Throughout the paper, we adopt a distance of 7.1 $\pm$ 0.3~kpc as the distance to BH~261.

The $\rm [Fe/H]=-$1.25 \citet{dekany21} metallicity for the BH~261 RR Lyrae star is based on the 
\citet{for11, chadid17, sneden17, crestani21} metallicity scale, abbreviated CFCS.
This is different than the $\rm SP\_ACE$ and 
APOGEE metallicity scale.  To quantify the difference between 
these two metallicity scales, the average \citet{dekany21} metallicity of RR Lyrae stars 
in the bulge GCs with at least 6 RR Lyrae stars is determined.  Two of those GCs have 
published $\rm [Fe/H]$ abundances from APOGEE's ASPCAP.  NGC~6642 has an 
APOGEE derived $\rm [Fe/H] = -$1.11 \citep{geisler21} and 19 OGLE RR Lyrae stars 
with an average CFCS photometric $\rm [Fe/H] = -$1.42.  FSR1758 has an APOGEE 
derived $\rm [Fe/H] = -$1.43 \citep{romerocolmenares21} and 9 OGLE 
RR Lyrae stars with an average CFCS photometric $\rm [Fe/H] = -$1.84.  
Therefore, the RR Lyrae star photometric metallicities are $\sim -$0.3~dex more metal-poor than 
the APOGEE ASPCAP $\rm [Fe/H]$ metallicity.  The BH~261 RR Lyrae metallicity of 
$\rm [Fe/H]=-$1.25 corresponds to $\rm [Fe/H]= -$0.95~dex on the 
APOGEE/ASPCAP scale, in agreement with the 
SP\_ACE derived metallicity of BH~261 giants in \ref{sec:feh}.  

\subsection{Color-magnitude diagram}
The BDBS photometry combined with {\it Gaia} astrometry allow a modern optical CMD of BH~261 to be 
constructed and so we focus first on the cluster itself in an effort to validate the cluster parameters determined independently without the use of BDBS photometry ($e.g.,$ distance and metallicity).  
Figure~\ref{cmd} shows the de-reddened proper-motion cleaned $u_0$ versus $(u-g)_0$ and 
$r_0$ versus $(r-z)_0$ CMD of a region within 3.2' of BH~261.  
A 3.2' radius is chosen because it is large enough to encompass both enough cluster and field BDBS stars to see a differentiation between the two when separated by proper motion, and it is also small enough where most of the radial velocity confirmed cluster stars are present (see Figure~\ref{GC_RV_zoom}).  
All stars have been dereddened using the extinctions from the \citet{simion17} reddening map, which has a resolution of 1' x 1'.  
The reddening vectors were computed using \citet{green18} for the $grizy$ bands and \citet{schlafly11} for the $u$ band, as outlined in \citet{johnson20}. The \citet{green18} extinction vector is preferred as it is based on a combination of broad band stellar colors and APOGEE spectra, where most of the APOGEE reference stars used belong to the disk and bulge.   \citet{kader23} derive high-resolution extinction maps for 14 GCs in BDBS and show that these reddening maps are in agreement with the VVV map used here. 
Still, we note the $u$-band extinction vector is notoriously difficult to calibrate, and large uncertainties in the $u$-band extinction can arise from small variations in the reddening law between different lines-of-sight.  
The range of extinction values within the central 3.2' of BH~261 varies from 
$E(B-V) \sim$ = 0.25 - 0.36, with a mean of $E(B-V) =$0.29~mag.   

The isochrones used are from the publicly available Modules for Experiments in 
Astrophysics (MESA) Isochrones and Stellar Tracks (MIST) database \citep{choi16}.  
The isochrones were transformed from theoretical coordinates to the appropriate bandpasses using a combination of the SDSS and PanSTARRS color transformation schemes. The $\alpha$-element enhancement is accounted for following the procedure described in \citet{joyce23}.
The shorter wavelength passbands ($e.g.,$ $u$-band) allow for 
the largest discrimination between isochrones with $e.g.,$ different metallicities and ages, but 
shorter wavelength passbands are also more sensitive to reddening and extinction variations, {as discussed above}.  
%The \citet{simion17} reddening map has a resolution of 1', but in the inner Galaxy, reddening variations down to a few arcseconds have been seen in maps of specific windows \citep[e.g.,][]{gosling09, gonzalez12, surot20, kader23}. 

The isochrones with metallicities of $\rm [Fe/H] \sim -$0.9 to  $\rm [Fe/H] \sim -$1.1~dex with an old age ($\sim$13-13.5~Gyr) are in agreement with the $u-g$ CMD.  
A 13.4~Gyr isochrone age was used to be similar to the 13.4$\pm$1~Gyr estimated by \citet{barbuy21}.  
Most bulge GCs with $\rm [Fe/H] \sim -$1.0 and BHB have ages in this 
range \citep[e.g.,][]{kerber18}, and it has been shown that one avenue to produce such 
metal-rich GCs with a BHB is by them having a very old age \citep[e.g.,][]{lee94}.  
%This is because theoretical HB models show that a cluster's HB becomes redder with increasing 
%$\rm [Fe/H]$ for a fixed mean mass-loss, HB helium abundance and cluster age.  
%In contrast, clusters will have bluer HBs with increasing age for a fixed metallicity, mean mass-loss and HB helium abundance.  Therefore, older ages for metal-rich clusters will 
%produce stars that populate the blue end of the HB \citep[e.g.,][]{stetson99, catelan01}.  
%Variations in He abundance, CNO abundance and core 
%mass could also lead to blue HBs in metal-rich GCs, but it has been postulated that these 
%effects are inconsistent with the observed properties of bulge main-sequence turnoffs and the 
%properties of the RR Lyrae variables \citep{lee94, lee16}. 
%
We note that very low metallicities are not needed for stellar relics in the bulge -- 
the chemical enrichment of the bulge is faster than many other places in the 
MW \citep[e.g.,][]{zoccali06, bensby13}, and a flat age-metallicity relation for inner 
Galaxy GCs has been established \citep{marinfranch09, massari19}.

The radial velocity members falling along the BHB of the cluster (see Figure~\ref{cmd})
confirm that BH~261 does have an extended BHB, despite it being relatively metal-rich.  
It has been suggested that BH~261's broad HB could be due to a number of blue straggler (BS) stars.  
Although this may be the case, we find that the contamination in this region of the CMD from field stars in not trivial.  
From our sample of 5 spectroscopically targeted possible BHB stars (all with 
proper motions consistent with BH~261), 2/5 have radial velocities 
excluding them from being cluster members.  This ratio is similar to the giant stars we targeted and 
highlights the difficulty obtaining clean cluster samples from proper motions and position on the CMD alone.

The $ugrizY$ BDBS photometry of all stars within 3.2' of BH~261 is presented in 
Table~\ref{tab:bdbs}.  

\subsection{Velocities}
The derived radial velocities as a function of distance from the cluster center are shown in 
Figure~\ref{GC_RV_zoom} (left panel).  
There is a grouping of stars within 4' of the cluster with radial velocities between 
$-$35~km~s$^{-1}$ and $-$80~km~s$^{-1}$ which we consider the most probable 
member stars currently within BH~261.  To search for systematic offsets between different 
samples of BH~261 stars, our stars are cross-matched with the sample presented in \citet{geisler23} and 
\citet{barbuy21}.  There are two stars in our sample that overlap with those in \citet{geisler23} --
{\it Gaia}-4050600806719928576 and {\it Gaia}-4050624308743727744.  
The radial velocities presented here agree 
within one-sigma of the radial velocities reported in Table~2 of \citet{geisler23}, both when 
the total velocity error of 7.5~km~s$^{-1}$ is adopted for the \citet{geisler23} measurements, 
(which arises from the error in centering the image in the spectrograph combined with 
the standard deviation of the different cross-correlations) as well as when the smaller, 
statistical errors in velocity of $\sim$2~km~s$^{-1}$ are adopted.  

The most probable cluster members of BH~261 are listed in Table~2, along with 
(1) the {\it Gaia} DR3 ID of each star, 
(2) the right ascension of the star from {\it Gaia}, 
(3) the declination of the star from {\it Gaia} in degrees, 
(4) the proper motion in right ascension direction as provided, 
by {\it Gaia} ($\rm \mu_{\alpha}* = \mu_\alpha cos \delta$), 
(5) the proper motion in declination as provided by {\it Gaia},  
(6) the heliocentric radial velocity, 
(7) the $\rm [Fe/H]$ metallicity from SP\_ACE, and 
(8) the distance the star is from the cluster center. 

The $\rm [Fe/H]$ metallicities of these stars, as discussed in \ref{sec:feh} below, are also consistent with 
being more metal-poor than the field population.  Therefore, the stars most likely currently within the cluster BH~261 (1) are within the cluster tidal radius, 
(2) have an RV that falls within the error 
plus intrinsic dispersion (generously adopted as $\pm$15 km~s$^{-1}$) from the cluster mean, 
(3) have an $\rm [Fe/H]$ value within $\pm$0.3~dex of the mean metallicity of the cluster, 
and (4) have a proper motion that lies within two standard 
deviations from the cluster mean.  These criteria have been used by a number 
of similar studies discriminating between bulge cluster 
members and surrounding field stars \citep[e.g.,][]{parisi22, dias22, geisler23}. 

The mean velocity of these 12 stars is $\rm <RV>$ = $-$56$\pm$3~km~s$^{-1}$ 
with a radial velocity dispersion of $\rm <\sigma>$ = 7.0$\pm$1.9~km~s$^{-1}$. This velocity dispersion is 
higher than reported in previous studies, but is also based on a sample size that is a factor of 4 larger.  
Although most foreground stars should be distinguishable with parallax, 
background stars can not.  Removing the star with the most negative radial velocity, 
a star 1.9 arcminutes from the 
cluster center,  a mean velocity of  $\rm <RV>$ = $-$54$\pm$2~km~s$^{-1}$ 
with a radial velocity dispersion of $\rm <\sigma>$ = 5.0$\pm$1.7~km~s$^{-1}$ is found.  This 
brings down the velocity dispersion.  Note that this star has a $\rm [Fe/H]$ metallicity 
consistent with the metallicity of BH~261, and is more metal-poor than the field, which is why it 
is still included as a potential cluster member.  
Removing the horizontal branch stars, which have larger radial velocity uncertainties, 
as well as one giant with a radial velocity uncertainty of 10~km~s$^{-1}$, the 
mean velocity is $\rm <RV>$ = $-$61$\pm$2.6~km~s$^{-1}$ 
with a radial velocity dispersion of $\rm <\sigma>$ = 6.1$\pm$1.9~km~s$^{-1}$.

Including the other four independent radial velocity measurements -- 3 stars from 
\citet{barbuy21} and 1 star from \citet{geisler23} -- gives a sample of 16 stars.  
We then remove the stars with both 
the highest and lowest velocities to obtain a mean velocity of 
$-$53.6$\pm$2.0~km~s$^{-1}$ with a dispersion of 5.9$\pm$1.9~km~s$^{-1}$.  

There may be other stars belonging to BH~261, as stars with radial velocities in this range exist 
out to as far as our observations go.  Before evaluating the likelihood of extra-tidal stars 
around BH~261, $\rm [Fe/H]$ metallicities are calculated.  

\begin{figure}
\centering
\mbox{\subfigure{\includegraphics[width=8.2cm]{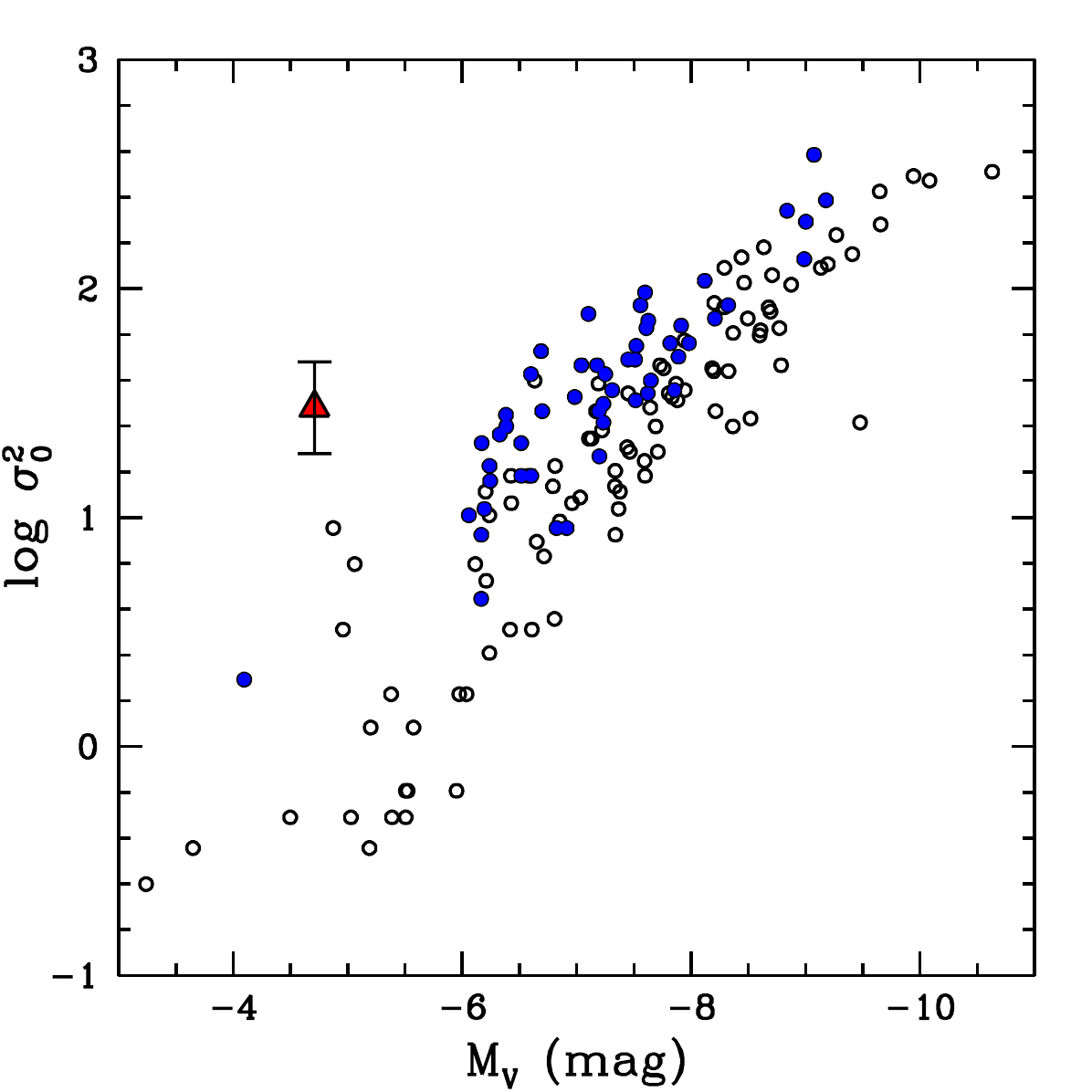}}}
\caption{Absolute integrated magnitude ($\rm M_V$) and velocity dispersion (log $\sigma_0^2$) for 
clusters analyzed by \citet{baumgardt18}. 
The open black points indicate globular clusters that are further than 3.35~kpc from the 
Galactic center, where as the filled blue points indicate inner Galaxy globular clusters -- those with galactocentric 
distances less than 3.35~kpc.  The red triangle designates BH 261.
}
\label{mv_logsig}
\end{figure}

Figure~\ref{mv_logsig} shows the Milky Way GCs analyzed in \citet{baumgardt18} 
with measured absolute magnitudes and intrinsic velocity dispersion values.  
The $\sigma_0^2$ parameter comes from the following equation 
$\sigma_0^2$ = $\sigma_{\rm vel}^2$ - $\sigma_{\rm errors}^2$, where $\sigma_{\rm vel}$ is the 
standard deviation of the radial velocity distribution of the cluster members 
and $\sigma_{\rm errors}$ is the mean error of the velocity measurements.
Using the radial velocity measurements from the 11 red clump and giant stars\footnote{We neglect 
the RR Lyrae star due to its large radial velocity uncertainty}, 
a $\rm log~\sigma_0^2$ = 1.7~km~s$^{-1}$ is found.  Removing the two stars 
in the sample with the highest and lowest radial velocities gives a $\rm log~\sigma_0^2$ = 1.3~km~s$^{-1}$.

A velocity dispersion value based on the spread in the proper motions only 
can also be calculated.  Because the proper motions are derived independently 
than the radial velocities presented here, this could be a further check as to the validity 
or our radial velocities.  
%This can be useful in evaluating how/if the $\rm log~\sigma_0^2$ value changes by culling the sample in radial velocity space in  an effort to clean potential contaminants.  
In this case, the {\it Gaia} proper motions are converted into tangential 
velocities ($v_t$) in km~s$^{-1}$ using the d=7100~pc distance found here and the relation 
$\rm v_t = 4.74\mu d$, where $\mu$ is the total proper motion in arcsec~yr$^{-1}$ 
and 4.74 is the conversion of distance (pc to km), angle (from arcsec to radians), and 
time (from years to seconds).
We similarly recover a $\rm log~\sigma_0^2$ = 1.5~km~s$^{-1}$.
%Removing the two stars with the highest and lowest radial velocities gives an identical 
%$\rm log~\sigma_0^2$ = 1.5~km~s$^{-1}$.  
If the radial velocity membership criteria 
adopted here is too generous, the radial velocity outliers do not significantly 
affect the $\rm log~\sigma_0^2$ value of the cluster, as supported by the independent 
$\sigma_0$ value determined from proper motion measurements.  

Figure~\ref{mv_logsig} indicates that inner Galaxy GCs typically have larger internal velocity 
dispersions at the same luminosity as compared to GCs in the halo and disk.  
BH~261 is anomalous in that it has an internal velocity dispersion that 
generally is in line for clusters with brighter intrinsic magnitudes.  
The paucity of bulge GCs with $M_V>-$6 is likely due to the difficulty of 
detecting and studying low luminosity GCs in the crowded and heavily-extincted bulge.  The 
analysis presented here improves the properties of BH~261, especially important in the 
low-luminosity GC regime.  

\subsection{$\rm [Fe/H]$ Metallicities}\label{sec:feh}
The near-infrared region around the CaT is ideal for the determinations of radial velocities due 
to the strong CaT lines.  There is also spectral information in this regime that can be used 
to constrain temperature, gravity and chemical abundances \citep[e.g.,][]{ruchti10, koch17}.
The SP\_ACE code \citep{boeche21, boeche16} was designed to derive stellar parameters 
and chemical abundances over the spectral resolution interval R = 2000-40,000 and 
over the wavelength interval 4800-6860 \AA~and over 8400-8924 \AA.  
It was originally developed for elemental abundance determination for the Radial Velocity Experiment 
(RAVE) \citep[][]{kunder17, steinmetz20}, which covers the same wavelength range as the 
spectra collected here.  Although SP\_ACE can be used 
to measure individual abundances for different chemical species (Mg, Al, Si, Ca, Ti, Fe, and Ni), 
only $\rm [Fe/H]$ abundances are presented here.  The lower signal-to-noise of the 
spectra give the most reliable results for the Fe I and Fe II lines, which are the most numerous in 
our wavelength regime.
SP\_ACE was fitted to the 8450-8493\AA, 8503-8535\AA, 8550-8660\AA~and 
8670-8800\AA~wavelength regimes in order to avoid the strong CaT lines which often 
cause difficulties in precise abundances and stellar parameter determination.

\begin{figure}
\centering
{\subfigure{\includegraphics[height=8.5cm]{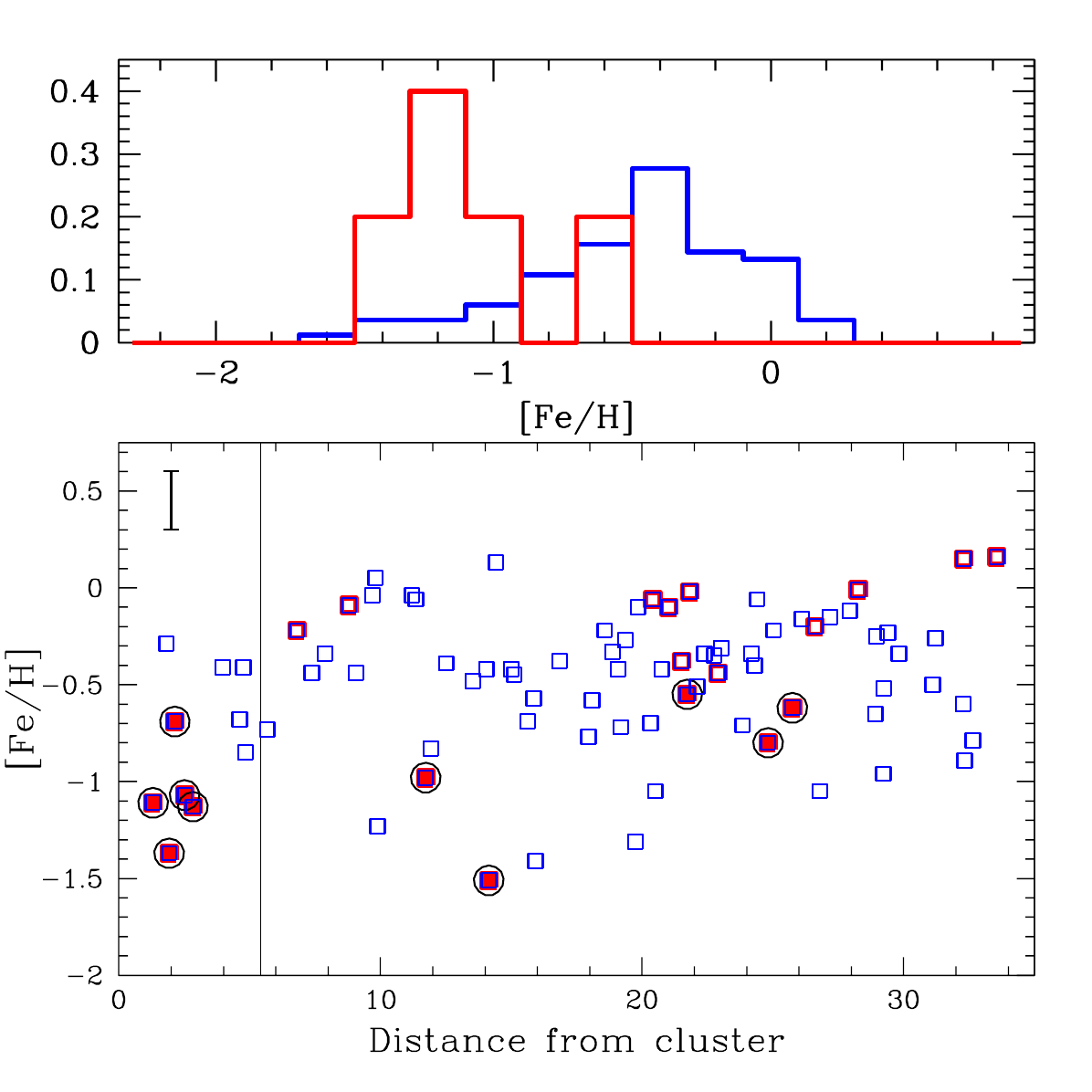}}}
%{\subfigure{\includegraphics[height=8.3cm]{graphs4Kunder_Jun12.pdf}}}
{\subfigure{\includegraphics[height=9.6cm]{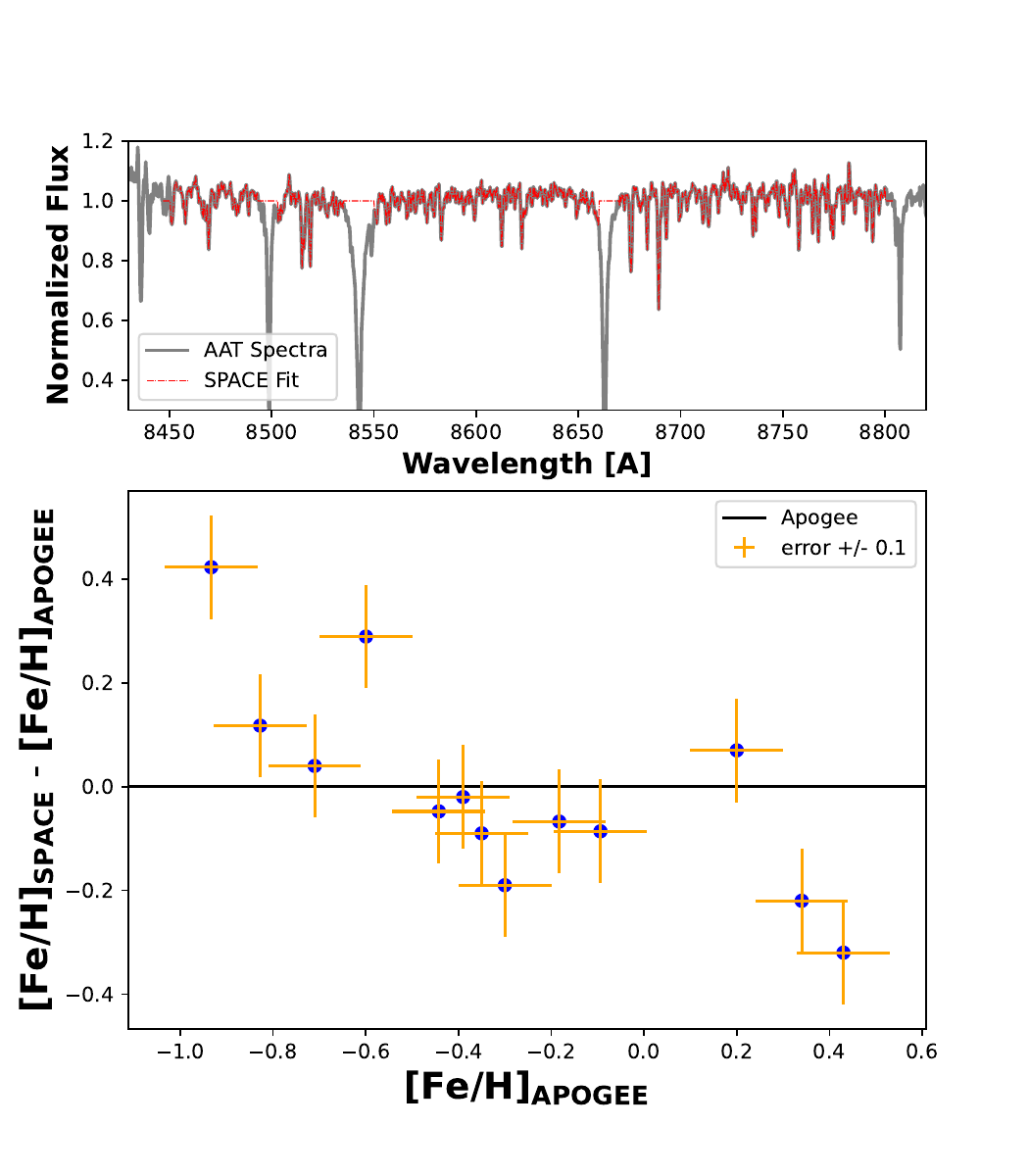}}}
\caption{
{\it Left:} A histogram of the $\rm [Fe/H]$ metallicity of our targeted stars as determined from 
$\rm SP\_ACE$ is shown in blue.  
Five giant stars within the cluster tidal radius with 
radial velocities consistent with the cluster are high-lighted in red.  There are five 
giants with both $\rm [Fe/H]$ metallicities and radial velocities outside of the tidal radius; 
these are extra-tidal giant star candidates.  The tidal radius determined here (\ref{sec:et}) is indicated 
by the solid line at a distance of 5.4' from the cluster center.  
{\it Right Top:}  
The observed sample spectra (black) of bh261\_1\_197.fits of the BH~261 giants within the tidal radius of the cluster 
is shown.  The best-fit spectra from $\rm SP\_ACE$ is overlaid in red, where only the portion of the spectra 
sampled by $\rm SP\_ACE$ is shown.  Note that the Campos wavelength axis used here 
extends to a longer wavelength range than used in \citet{koch17}.
{\it Right Bottom:}  
A comparison between seven APOGEE bulge giants and the $\rm [Fe/H]$ derived from $\rm SP\_aCE$ using the Campos wavelength range.
}
\label{GC_Fe}
\end{figure}

The wide coverage of APOGEE stars in the inner Galaxy allowed us to 
allocate a few fibers in our science fields to the re-observation of APOGEE bulge giants. 
The higher resolution of the APOGEE spectra (R$\sim$22,000 vs our 
R$\sim$10,000 spectra) as well as the high SNR for APOGEE stars ($\sim$100-200), 
made it advantageous to use these as calibration standards.
In total, 9 bulge giants in the APOGEE survey were observed during our run with our particular setup; 
they all have measured stellar parameters released in DR17 and span a wide range of $\rm [Fe/H]$ 
metallicities (see Figure~\ref{GC_Fe}).  
We also incorporated 6 APOGEE stars that were observed in previous AAT runs by our group with the same AAOmega setup to use as calibration standards. 

Figure~\ref{GC_Fe} (right panel) shows the SP\_ACE $\rm [Fe/H]$ metallicities as 
compared to those published by APOGEE.  The temperature and gravity regime of SP\_ACE is 
within $\rm T_{eff}$=[3600,7400]~K and log~$g$=[0.2,5.0].  For the two APOGEE 
stars with temperatures that are a few hundred Kelvin cooler than 3600~K, SP\_ACE did not 
converge, and therefore not all APOGEE stars observed were able to serve as metallicity 
standards.  
The uncertainty in $\rm [Fe/H]$ from SP\_ACE as compared to the APOGEE metalliciites 
is $\sim$0.2~dex.  There is an indication that SP\_ACE under-predicts the $\rm [Fe/H]$ for 
high-metallicity stars, and over-predicts the $\rm [Fe/H]$ in the low-metallicity regime, 
but SP\_ACE is able to reproduce the $\rm [Fe/H]$ of our observed spectra between the 
range of $\sim -$0.9 to $\sim$ +0.2~dex.  

\begin{figure*}
\centering
\mbox{\subfigure{\includegraphics[width=12.2cm]{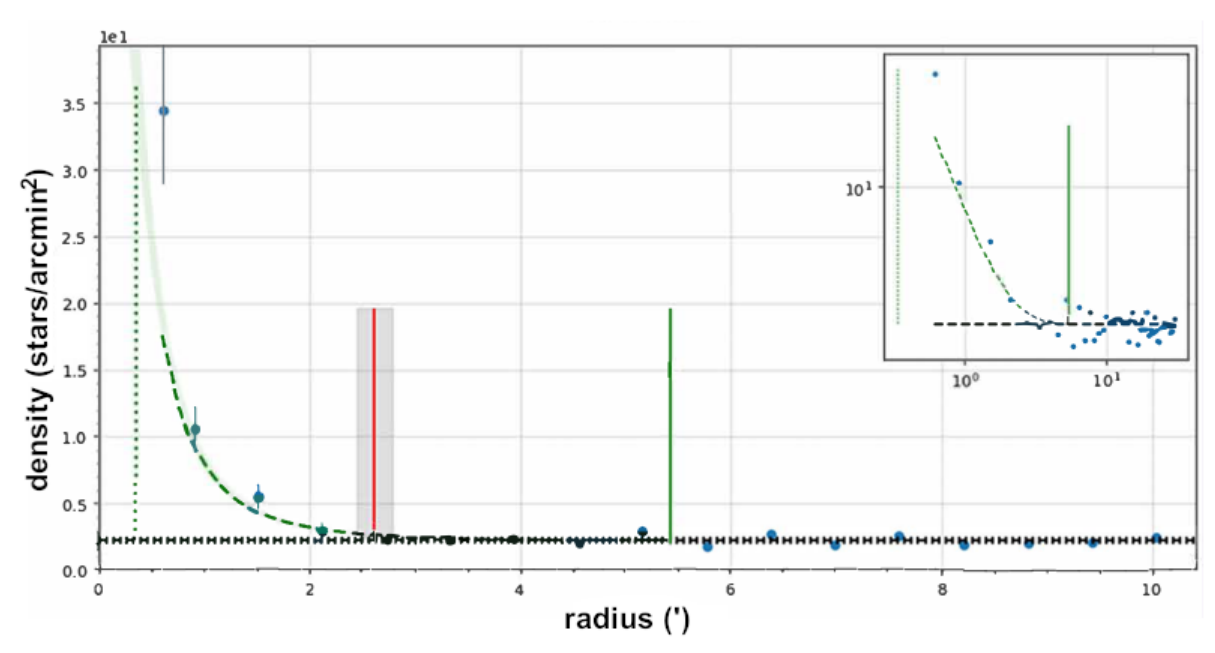}}}
\caption{
The combined BDBS photometry with $Gaia$ proper motion radial density profile of the BH~261 
cluster region. The dots are the stars per arcmin$^2$ taking the cluster center as the origin.  
The horizontal dashed black line indicates the field density.  
The King profile fit is indicated with the green dashed curve and the cluster core radius 
is indicated by the dotted vertical green line.
The red vertical line indicates the assigned radius of BH~261 with the uncertainty region 
marked as a gray shaded area.  The tidal radius is indicated by the solid vertical line 
at radius of 5.38 arcminutes.  A rescale of the main plot is shown in the top inset.
}
\label{asteca}
\end{figure*}

None of the 3 observed BHB stars have SP\_ACE metallicities 
that could be measured, since BHB have temperatures hotter than $\sim$8000~K.  
Also, hot BHB stars can have atmospheric effects like levitation and diffusion that mask 
their true abundances.  SP\_ACE did converge to 
provide a $\rm [Fe/H]$ estimate for 5 of the 8 non-calibrating giants observed.  The other three giants have a low 
SNR which was the likely why SP\_ACE failed to provide metallicities for these stars.  A weighted-
mean metallicity of 
$\rm < [Fe/H] > = -$1.07 $\pm$ 0.22 is found.  No evidence of a spread in the metallicity in 
BH~261 is seen, but our sample size is small and our formal $\rm [Fe/H]$ uncertainty is 0.2~dex.

\subsection{Extra-tidal stars}\label{sec:et}
In order to find signatures of BH~261 dissolving by the strong tidal field of the Milky Way, 
the tidal radius of the cluster needs to be known.  
Unfortunately the tidal radius of BH~261 is uncertain.  There are difficulties in defining the 
tidal radius, both theoretically and observationally, and although there are correlations between 
the calculated theoretical and observational radii, it is not uncommon to find 
discrepancies between tidal radii estimates when using theoretical and observational 
approaches \citep[e.g.,][]{moreno14}.  \citet{ortolani06} find the density profile 
merges with the background at 3.4$\pm$0.4~arcminutes, and the 2010 edition of the 
\citet{harris96} catalog list a tidal radius of 4' for BH~261.  The tidal radius listed in 
\citet{baumgardt21} is 20.63 pc, which at the distance listed 
in their catalog (6100 pc), corresponds to 11.6'.  
% type into google:  20.63/6100 radians to arcmin

We use ASteCA (Automated Stellar Cluster Analysis) with the BDBS photometry combined with 
$Gaia$ astrometry in an attempt to obtain an estimate of the tidal radius of BH~261.  
ASteCA is a python code \citep{perren15} 
designed to perform a thorough analysis of star clusters (open or globular), modeling 
spatial, structural, and photometric parameters.  ASteCA can determine cluster membership probabilities by utilizing a decontamination algorithm. It allows estimation of the center and 
radius of the cluster, along with density profiles, luminosity functions, and color-magnitude 
diagrams to study the stellar population within the cluster.  

ASteCA was fed 7215 stars from BDBS with useful photometry and 
$Gaia$ astrometry and that are within 30 arcminutes from the center of BH~261.  
The sample of stars was constrained to have 
0.0~mas~yr$^{-1} < \mu_\alpha <$+5.0~mas~yr$^{-1}$ and 
0.0~mas~yr$^{-1} > \mu_\delta > -$5.0~mas~yr$^{-1}$, as well 
as parallax $<$0.4~mas, in an attempt to minimize field star contamination. 
In total, ASteCA was run $\sim$30 times, utilizing different BDBS and $Gaia$ color-combination 
CMDs, with tightening proper motion limits.
The solutions for the physical parameters derived from the King models \citep{king62, king66} 
remained stable once we tightened the proper motion limits. We used the extinction corrections from 
\citet{simion17}.

Limiting the proper motion to +2.0$< \mu_\alpha <$+5.0 
and $-$2.5 $>\mu_\delta>-$4.5, produced 5,596 BDBS stars, and
does not change the ASteCA determined cluster parameters significantly. 
The ASteCA King model-fit for an example of the BDBS photometry (with Gaia proper motions) is 
shown in Figure~\ref{asteca}, which produced an isochrone fit for the $u_0$ vs. $\rm (u-z)_0$,  
and returned the core, cluster, and tidal radii as $r_c =$0.34$^{+ 0.44}_{-0.28}$’, 
$r_{cl} =$ 2.60$^{+ 2.79}_{-2.45}$’, and $r_t =$ 5.42$^{+ 6.93}_{-4.04}$’.  
Using the latter proper motion limits, the averages for the 6 final runs were 
$r_c =$0.352$\pm$0.004', $r_{cl} =$2.58$\pm$ 0.001', and $r_t =$5.377$\pm$0.080'.  
%The cluster field is heavily contaminated with 
%non-members, but the cluster and tidal radii converges.
%ASteCA’s King model fit returned 

%The individual uncertainties on the radii are listed first, and the
%uncertainties in parenthesis are the standard deviations after averaging the 
%means of all the colors. 
ASteCA is not the ideal tool for discerning extra-tidal structures, as the 
decontamination procedure uses the field stars outside the cluster radius to 
determine the cluster membership. However, the code’s unbiassed method of 
determining cluster parameters is useful to defining the radial density profile 
of the cluster and its members, thereby giving an indication of the tidal radius 
of the cluster.

\begin{figure}
\centering
\mbox{\subfigure{\includegraphics[width=8.2cm]{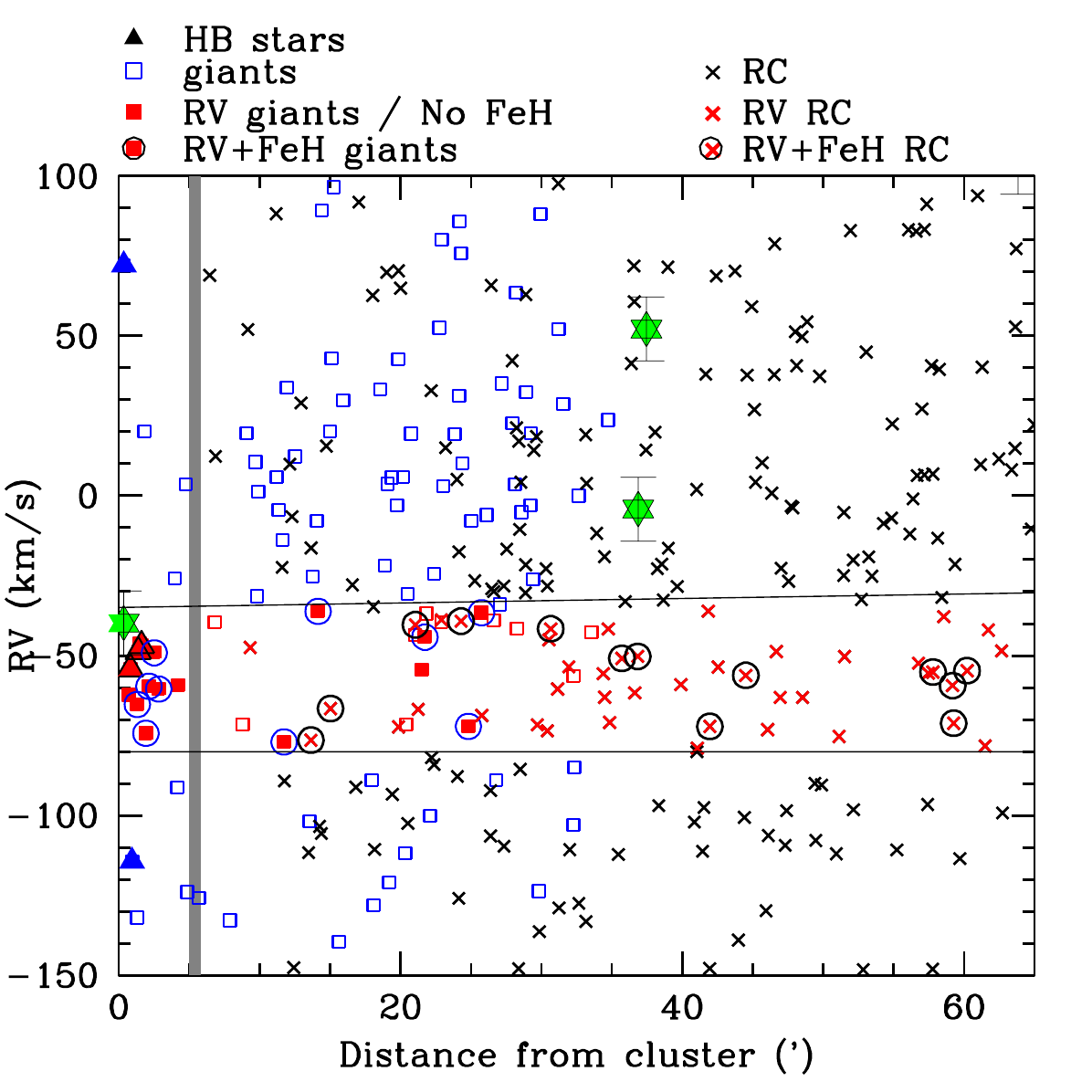}}}
{\subfigure{\includegraphics[height=8.2cm]{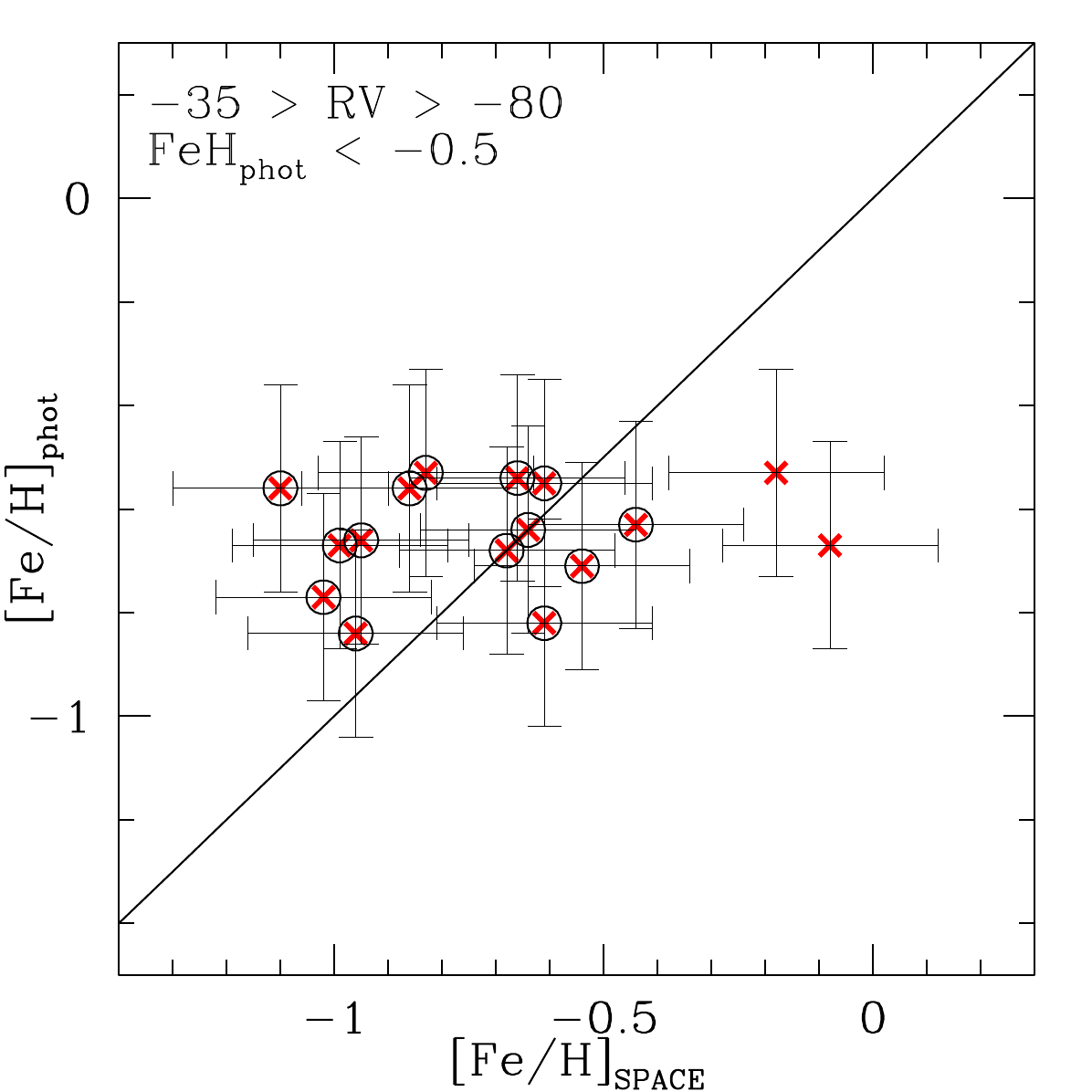}}}
\caption{
{\it Left:} The heliocentric velocities of our targeted stars within 65 arc-minutes of 
BH~261, with the tidal radius at $r_t$=5.4' indicated by the solid grey line.  
The uncertainties in radial velocity is $\sim$4~km~s$^{-1}$, except for the 
horizontal branch stars, where the RV uncertainty is $\sim$9~km~s$^{-1}$.  
Large (green) stars represent the targeted RR Lyrae stars.  
The stars with both SP\_ACE metallicities, photometric metallicities and radial velocities 
consistent with BH~261 are circled, and are potential extra-tidal stars belonging to BH~261.  
{\it Right:}  A comparison between the photometric $\rm [Fe/H]$ and spectroscopic $\rm [Fe/H]$ 
is shown for the stars with both photometric metallicities and radial velocities 
consistent with BH~261.  The two red clump stars with spectroscopic $\rm [Fe/H]$ metallicities 
that are discrepant from the photometric metallicities also have $T_{eff}$ and log~$g$ values 
suggesting they are not red clump giants.
}
\label{GC_RV_RC}
\end{figure}

We search for extra-tidal stars by targetting red clump stars as well as giants with 
proper motions consistent with BH~261.  
Figure~\ref{GC_RV_RC} (left panel) shows the velocities of all giants and red clump stars 
targeted spectroscopically.  These all have proper motions consistent with BH~261 (see 
Figure~\ref{pms}) and reach to 65 arc minutes from the center of the cluster.  This corresponds 
to approximately 6 - 10 times the cluster's tidal radius, depending on the exact calculation used 
for the tidal radius.  

Our observations 
detect BH~261 stars with distances out to $\sim$4~arcminutes from the cluster center, 
but beyond this distance, there is no clear over-density of stars with radial velocities 
consistent with BH~261.  There are a few giants and red clump 
stars with velocities similar to BH~261 between 5' and 10' from the cluster center, but 
these have metallicities that are more in-line with the bulge field as opposed to a GC.  
Beyond a distance of 10' from the cluster center, a handful of stars are identified that 
have both radial velocities and $\rm [Fe/H]$ metallicities consistent with BH~261.  
The most likely extra-tidal star candidates have both radial velocities 
and SP\_ACE metallicities consistent with BH~261; these are listed in Table~3.
%These are our most-likely extra-tidal candidates.  

It is estimated that contamination rates in the BDBS red clump star catalog is $\sim$30\% 
\citep[30\% of the stars actually are not red clump stars, instead belonging to $i.e.,$ the bulge red 
giant branch, inner disc or halo,][]{johnson22}.  Photometric metallicities may not be correct unless the star is 
on the red clump.  
In an attempt to remove any non-red clump members contaminating our sample as well as 
to confirm the $\rm [Fe/H]$ metallicities of the red clump stars, SP\_ACE is 
run on the red clump stellar spectra with both photometric metallicities and radial velocities 
consistent with BH~261, $i.e.,$ those star with radial velocities 
between $-$35 $>$ RV $>$ $-$80~km~s$^{-1}$ and a photometric $\rm [Fe/H] < -$0.5.  
These are the most probable extra-tidal stars.  Figure~\ref{GC_RV_RC} (right panel) shows a 
comparison between the photometric 
and spectroscopic $\rm [Fe/H]$ metallicities for the 16 red clump stars for which 
SP\_ACE converged and that have parameters indicating they could be extra-tidal 
stars.  Two of those have $\rm SP\_ACE$
metallicities that are too metal-rich to be part of the cluster, and these are excluded from the 
sample of potential extra-tidal star candidates.  

The Gala python package \citep{pricewhelan17, pricewhelan22} is used to generate a model of the dynamics of BH~261’s potential extra-tidal members. 
Gala provides several routines that allow the creation of mock stellar streams, by initializing new star particles at the cluster's Lagrange points with a specified frequency and with randomized velocity offsets consistent with a specified velocity dispersion.  
The orbits of each set of new star particles are then evolved forward in time within the combined gravitational potential of the cluster plus the potential of the Galaxy to reveal the spatial and kinematic structure that would be expected at the present day. 
Using the cluster’s position (RA = 273.527$^\circ$, Dec $-$28.635$^\circ$), distance (7.1 kpc), proper motion ($\mu_\alpha$ = 3.566 mas~yr$^{-1}$, $\mu_\delta$ = $-$3.590 mas~$^{-1}$), radial velocity ($-$61~km~s$^{-1}$) and mass (2.4x10$^4~M_\odot$), Gala calculates the cluster’s orbit in a Galactic potential. 
Here the adopted potential for the Milky Way is a three-component potential model consisting of the bar \citep[an implementation of the model used in][]{long92}, a Miyamoto-Nagai potential for the galactic disk \citep{miyamoto75}, and a spherical Navarro-Frenk-White \citep[NFW,][]{navarro97} potential for the dark matter distribution. The bar is tilted with respect to the x-axis by 25 degrees, has a mass 1/6 of the mass of the disk component and the long-axis scale length of the bar is set to 4~kpc \citep{blandhawthorn16}.  

We simulate the ejection of star particles using the Fardal Stream Generator \citep{fardal15}.  
The Fardal Stream Generator simulates the formation of extra-tidal structures via external tidal stripping, rather than more violent internal relaxation processes. 
The locations of the Lagrange points from which stars are ejected, as well as the velocity offsets the stars receive when they are ejected, are set to be consistent with the cluster's mass and density profile, and the velocity dispersion expected for a fully thermalized population.  
The frequency with which star particles are ejected, however, is non-physical: we simulate ejection events every 0.5 Myrs over the past 100 Myrs, to ensure that we densely sample all positions and velocities for which tidally stripped stars would be present. 
%We also project the cluster’s orbit backwards for 2 Gyr to explore the structure of older debris. 

\begin{figure}
\centering
{\subfigure{\includegraphics[width=5.9cm]{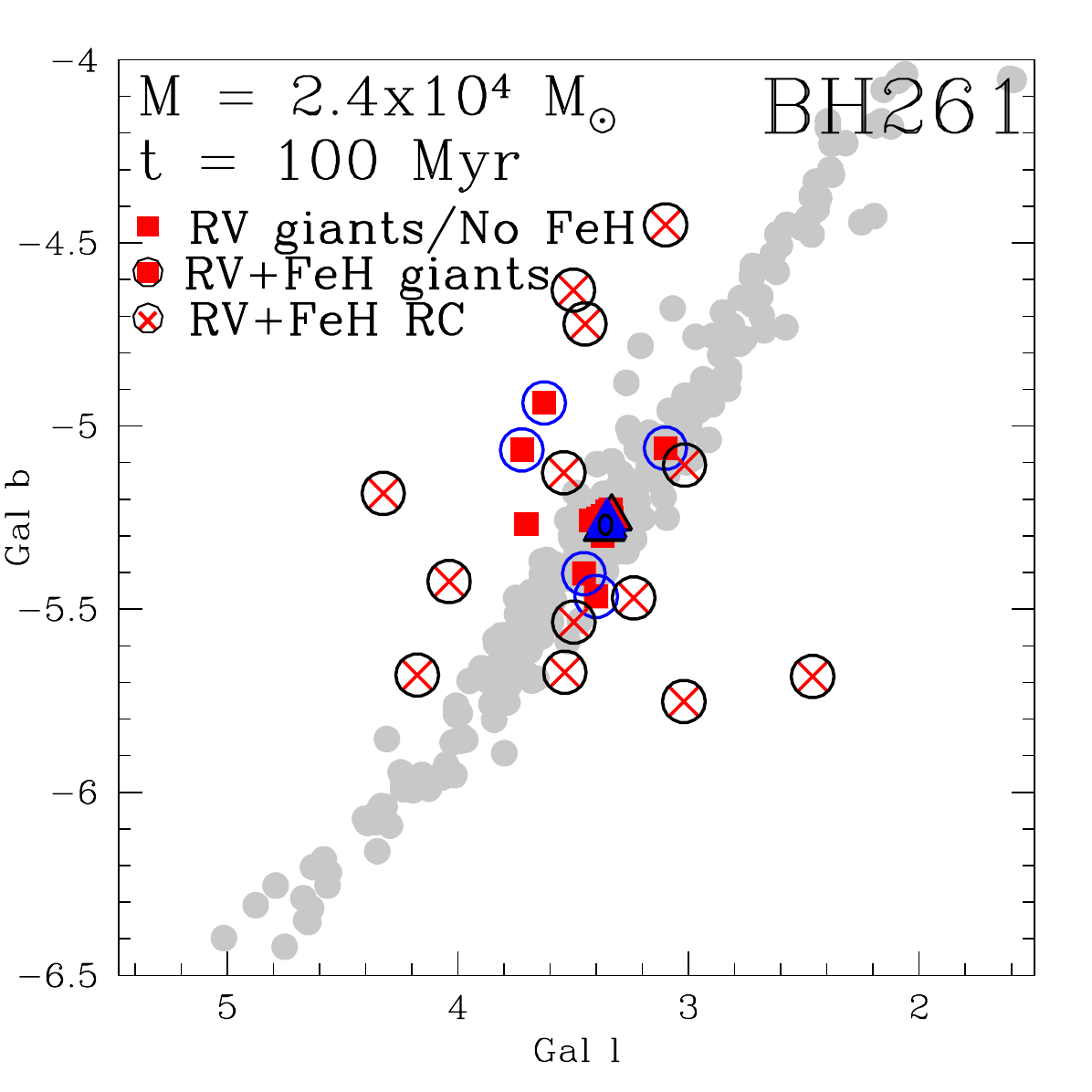}}}
{\subfigure{\includegraphics[width=5.9cm]{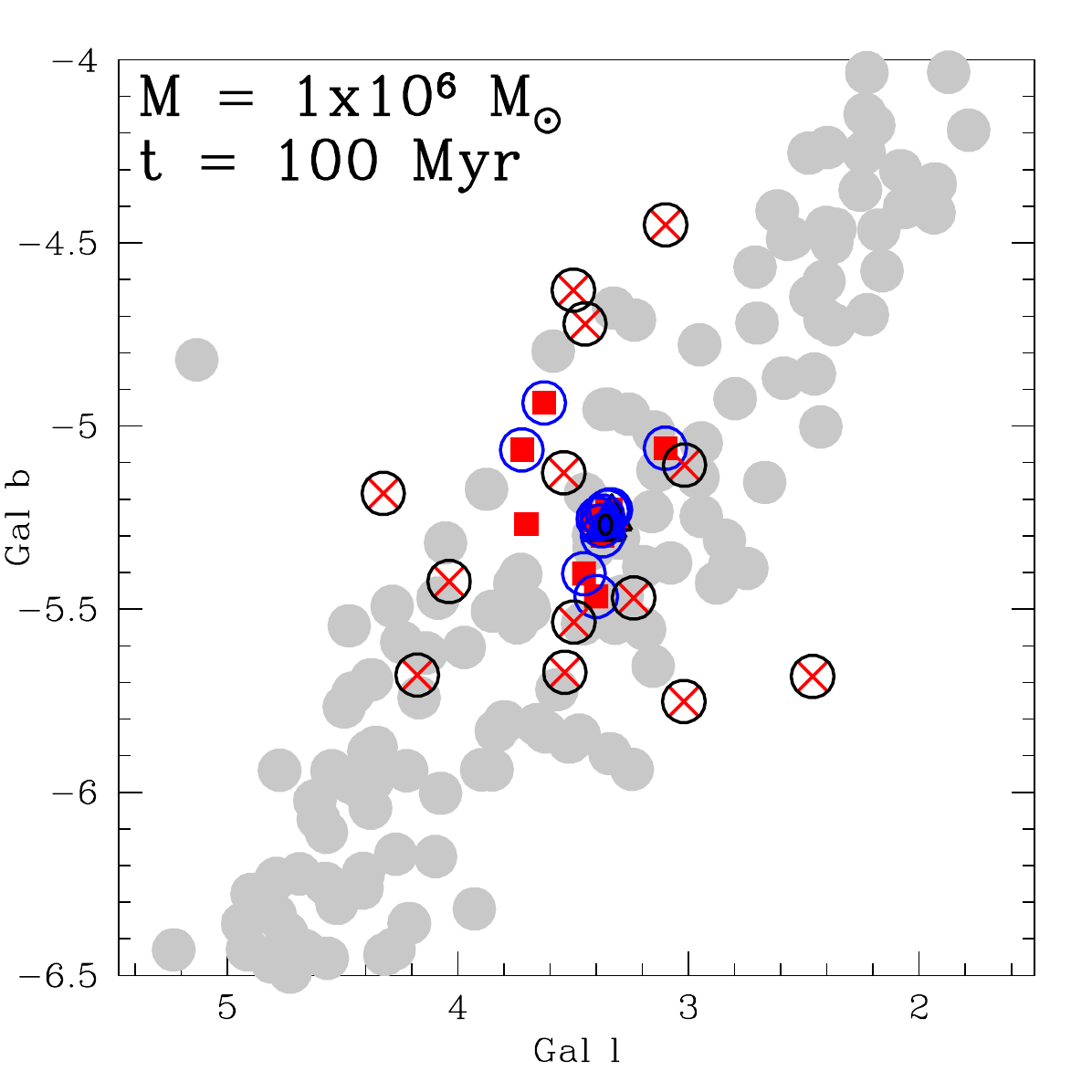}}}
\caption{The simulated tidal tails for BH~261 (grey), calculated with the cluster’s best fit parameters, 
are compared to the most promising BH~261 extra-tidal candidates presented here. 
The inset in the upper left corner designates the length of time the cluster orbit is integrated forwards 
as well as the mass of BH~261 used in the simulation.  The width of the stripped stars is 
tight when using a small GC mass ($i.e.,$ the mass inferred from its luminosity), and 
the simulated tidal tails occupy a greater area around the center of the cluster as a larger GC 
mass is used ($i.e.,$ the mass inferred from its velocity dispersion).  
}
\label{wwu_bh261_in}
\end{figure}

A comparison of the location of the candidate extra-tidal stars we have identified with 
the synthetic extra-tidal stars in the Gala simulations is shown in Figure~\ref{wwu_bh261_in} (left panel).  
Because the most recent stellar debris will be both physically closest to the cluster, and the most kinematically coherent, we focus on the debris produced within the last 100 Myrs, which is most amenable to detection in our spectroscopic observations. Also, the simulated debris produced within the last 100 Myrs is the least affected by the adopted potential of the Milky Way. 

The Gala simulation does suggest that ejected stars from BH~261 could show a stream of stars, with arms on the leading and trailing end of the orbital path of the cluster.  Our observations are limited to the brightest stars, which given the low-luminosity of BH~261, would likely not to have the spatial density to show a clear stream.  Still, we may be able to detect tidal disruption, which could follow a coherent structure.   
No stellar tidal streams have been seem emanating from bulge GCs to date, although a low-luminosty ($M_V = -$3.0 $\pm$ 0.5, similar to that found for the lowest-mass GCs) stream, the Ophiuchus stream, has been detected near the MW bulge region, above the center of the Galaxy \citep{bernard14}.  
Its old ($\sim$12~Gyr) and relatively homogeneously metal-poor population and $\alpha$-enhanced stars suggest that the progenitor would most likely be a globular cluster\citep[e.g.,][]{sesar15}.  However, because of its short length and short orbital period of the stream, it should have been disrupted fairly recently, but no progenitor is visible.  
In an attempt to explain the Ophiuchus stream, models and mechanisms to enhance the density of some stellar streams in the inner halo/bulge have been put forward, but to date no model can explain the short Ophiuchus stream with such a short orbital period \citep[e.g.,][]{hattori16, pricewhelan16, lane20}.  Observational analysis to detect additional substructures in the inner Galaxy is needed for a more complete understanding of the Milky Way's gravitational potential and therefore a better dynamical study of clusters as they pass through the inner Galaxy, which are complicated by also the influenced by the Galactic bar, it's rotations and how the bar has changed with time\citep[e.g.,][]{hattori16}.

Only a handful of the stars with radial velocities consistent with BH~261 are spatially coincident 
with the predicted 100~Myr tidal debris.  
Extra-tidal stars that do not fall along the cluster's current orbit could have arisen 
due to effects not included in Gala, such as stars ejected 
due to shocks caused by the tidal field of the Galaxy and/or stars ejected from 
tidal interactions with the galactic plane \citep[e.g.,][]{moreno14}, 
or a much larger dispersion of ejection velocities and angles due to intra-cluster 
interactions, such as those simulated by the core particle spray algorithm \citep{grondin23}.   
Although less significant, stars can also be ejected due to interactions with the giant 
molecular clouds \citep[e.g.,][]{amorisco16}, a process Gala is unable to incorporate.

Given the large velocity dispersion of BH~261 and hence a potentially larger dynamical mass, 
we also used Gala with a cluster mass more 
in-line with the dynamical mass of the cluster, $\rm 1x10^6 M_{\odot}$.  
Figure~\ref{wwu_bh261_in} (middle panel) shows that in this case, 
the comparison of the observed extra-tidal stars with simulations is improved.

High-resolution spectroscopy to chemically fingerprint the extra-tidal star candidates 
would allow a deeper characterization into the origin of these stars.  
%We note that stars stripped throughout the passage of BH~261 will 
%end up all over the inner Galaxy after several Gyrs.  
There are a handful of  
extra-tidal star candidates that fall along the predicted 
tidal debris, but it is likely that most of the candidate extra-tidal stars listed in Table~3 
are not recently stripped from the cluster.  
That few stars are currently being stripped from BH~261 would be in agreement with the cluster's 
low luminosity -- the cluster does not have as many stars left for it to lose today as it had 
in the past.  This would be consistent with 
BH~261 being an old, low-mass cluster so that any extra-tidal stars are likely 
to be white dwarfs, rather than more massive main sequence stars and giants.  In 
this case, it was be difficult to detect extra-tidal stars from our observations.  
If the extra-tidal candidates are confirmed to be bona fide 
extra-tidal stars, that might indicate the velocity dispersion of BH~261 is driven by 
tidal heating rather than a high mass-to-light ratio.

\section{Conclusions}
In order to better understand the bulge field population as well as to 
observe the processes inner Galaxy GCs undergo in their passage 
through the inner Galaxy, the MWBest spectroscopic survey is identifying stripped stars from 
inner Galaxy GCs.  
%Low-mass/low-luminosity globular clusters may be a bridge linking the bulge GC population to the 
%bulge metal-poor stellar component we observe today.  This is because 
On average, mass lost from low-mass/low-luminosity globular clusters such in the inner parts 
of the Milky Way, such as BH~261, 
will be considerably larger than for clusters with present-day masses larger 
than $\rm 10^5 M_\odot$ \citep{baumgardt21}.  These low-mass/low-luminosity clusters 
started off with masses greater than $\sim \rm 10^6 M_\odot$ and have lost mass as 
they move through the Galaxy.  By integrating the orbits of the Milky Way GC backward in 
time and applying suitable recipes to account for the effects of dynamical friction and mass-loss 
of stars to the clusters, \citet{baumgardt21} show that especially clusters inside the central 2~kpc 
of the MW have lost a large portion ($\sim$80 \%) of their initial populations.  
BH~261 has the smallest mass of the Milky Way GCs listed in \citet{baumgardt21}, 
weighing in at $\sim$2.4$\pm$0.6~x~$\rm10^4~M_\odot$.  
It is also one of the few low luminosity bulge GCs, with an absolute magnitude of 
$M_V$=$-$4.43 mag.  In fact, there are only two well-studied bulge GCs (those with more 
than 10 stars with radial velocity measurements) listed in the \citet{baumgardt21} 
catalogue with $M_V>-$6, the other one being ESO\_452-SC11 \citep[see][]{koch17, simpson17}.  

In order to carry out a search for extra-tidal stars around the BH~261, a better characterization 
of this GC is needed.  Using BDBS photometry combined with $Gaia$ astrometry, the radial density profile 
of the cluster region is decomtaminted and fit with a King profile using ASteCA.  
In this way we derive a core radius of BH~261 of $r_c =$ 0.35', 
a cluster radius of $r_{cl} =$ 2.6', and a tidal radius of $r_t =$ 5.4'.

We carried out the largest spectroscopic analysis of stars within the tidal radius of BH~261.  
From the 7 giant stars with the best radial velocities within $\sim$4' of the center of BH~261, a 
mean velocity of $\rm <RV>$ = $-$61$\pm$2.6~km~s$^{-1}$ 
with a radial velocity dispersion of $\rm <\sigma>$ = 6.1$\pm$1.9~km~s$^{-1}$ is found.  
When all 12 BH~261 stars within $\sim$4' of the center are considered a mean velocity 
of $\rm <RV>$ = $-$56$\pm$1.7~km~s$^{-1}$ 
with a radial velocity dispersion of $\rm <\sigma>$ = 7.0$\pm$1.9~km~s$^{-1}$ is found.  
This average velocity is consistent with that of \citet{barbuy21}, who find 
$\rm <RV>$ = $-$57.9$\pm$4.3~km~s$^{-1}$.  It differs from that of 
\citet{baumgardt19} who find $\rm <RV>$ = $-$29.4~km~s$^{-1}$, and it differs from that of 
\citet{geisler23} who find $\rm <RV>$ = $-$44.9$\pm$3.8~s$^{-1}$.  
However, the stars observed here encompass the velocity values reported in previous 
studies.  For example, the three observed stars in \citet{barbuy21} span a large 
velocity range with velocities of 
$-$67.65$\pm$3.65~km~s$^{-1}$, $-$57.93$\pm$4.28~km~s$^{-1}$ and 
$-$29.57$\pm$5.85~km~s$^{-1}$. 
Similarly, the three observed stars in \citet{geisler23} have velocities of 
$-$52.5$\pm$1.9~km~s$^{-1}$, $-$42.33$\pm$1.2~km~s$^{-1}$ and 
$-$39.9$\pm$2.3~km~s$^{-1}$.  
What is consistent in all spectroscopic studies of BH~261 to date is that the velocity spread is 
not insignificant.  The larger sample of stars presented here allows for a more robust value 
for a mean velocity.  

The SP\_ACE code was utilized for the determination of $\rm [Fe/H]$ metallicities, and from 
spectra of 5 giants, an average $\rm [Fe/H] \sim$ $-$1.1 $\pm$ 0.11~dex is found.  
By identifying a RR Lyrae star in BH~261, the distance to the cluster is found to be 
7.1$\pm$0.4~kpc.  This, as well as direct spectroscopic measurements of $\rm [Fe/H]$ 
from 5 giant stars in BH~261, confirms BH~261 is on the near-side of the bulge.  As discussed also 
in \citet{gran22}, this shorter distance indicates BH~261 has an abnormally low-luminosity as compared 
to its stellar velocity dispersion.  
New BDBS photometry in the $ugrizY$ passbands is presented of the central region of 
BH~261 and is used to check for consistency between the cluster parameters and the 
optical CMD of BH~261.  The MIST isochrones with the cluster's distance and metallicity derived 
in this work (7.1~kpc and $-$1.1~dex) show good agreement with the BDBS CMDs and with an 
old cluster age, of $\sim$13~Gyr.  Such an age is similar to other bulge GCs with blue HBs \citep{kerber18}.  

A search for candidate extra-tidal stars spanning the radial velocity and proper motion range 
of BH~261 was carried out.  A few of our most promising extra-tidal candidates -- those with 
radial velocities, proper motions and $\rm [Fe/H]$ metallicities consistent with BH~261 -- 
are consistent with Gala simulations of the dynamical evolution of the cluster using the present day mass 
of the cluster.  But most are only consistent with recent tidal debris from BH~261 if a larger cluster mass is used.  
BH~261 is an old, low-mass cluster, and it may be that most stripped stars today are white dwarfs, rather than more massive giants we are able to target spectroscopically. 

\begin{deluxetable*}{lllcccccccccccc}
\tabletypesize{\scriptsize}
\tablenum{1}
\tablecaption{Blanco DECam Bulge Survey (BDBS) photometry of stars within 3.2' from BH~261
\label{tab:bdbs}}
\tablehead{
\colhead{{\it Gaia} ID} & \colhead{RA (deg)} & \colhead{Dec (deg)} & \colhead{$u$} & \colhead{$u_{err}$} & \colhead{$g$} & \colhead{$g_{err}$} & \colhead{$r$} & \colhead{$r_{err}$} & \colhead{$i$} & \colhead{$i_{err}$} & \colhead{$z$} & \colhead{$z_{err}$} & \colhead{$y$} & \colhead{$y_{err}$}
}
%\decimalcolnumbers
%\decimals
\startdata
\hline
 4050647188105074639 & 273.46301 & $-$28.67483 & 21.931 & 0.018 & 20.106 & 0.019 & 19.503 & 0.03 & 19.061 & 0.023 & 18.966 & 0.038 & 18.937 & 0.009\\
4050647879579387008 & 273.46301 & $-$28.63367 & 20.758 & 0.009 & 18.013 & 0.021 & 16.918 & 0.004 & 16.532 & 0.001 & 16.295 & 0.001 & 16.096 & 0.003\\
4050647192336176640 & 273.46302 & $-$28.666 & NaN & NaN & 19.253 & 0.041 & 19.081 & 0.003 & 18.8 & NaN & 18.719 & 0.004 & 18.699 &0.036\\
4050648051378176128 & 273.46305 & $-$28.61794 & 21.007 & 0.016 & 19.4 & 0.007 & 18.645 & 0.005 & 18.434 & 0.016 & 18.263 & 0.046 & 18.333 & 0.009\\
4050648051330018176 & 273.46309 & $-$28.61518 & 18.898 & 0.012 & 16.4 & 0.008 & 15.408 & NaN & 14.963 & 0.011 & 14.611 & 0.031 & 14.448 & 0.051\\
\enddata
\end{deluxetable*}

\begin{deluxetable*}{lccccccc}
\tabletypesize{\scriptsize}
\tablenum{2}
\tablecaption{Positions, $Gaia$ proper motions, radial velocities and $\rm [Fe/H]$ metallicities 
of the probable member stars of BH~261
\label{tab:rvfe}}
\tablehead{
\colhead{{\it Gaia} ID} & \colhead{RA (deg)} & \colhead{Dec (deg)} & \colhead{$\mu_{\alpha}$ (mas~s$^{-1}$)} & \colhead{$\mu_{\delta}$ (mas~s$^{-1}$)} & \colhead{HRV (km~s$^{-1}$)} & \colhead{$\rm [Fe/H]$} & \colhead{r (') }
}
%\decimalcolnumbers
\startdata
\hline
 4050671244212595584 & 273.525876 & $-$28.601011 & 3.431 $\pm$ 0.084 & $-$3.410 $\pm$ 0.064 & $-$59.6 $\pm$ 2.9 & $-$0.69$\pm$0.2 & 2.14 \\
4050624029553084928 &  273.562146 & $-$28.638807 &  3.727 $\pm$ 0.052 & $-$3.612 $\pm$ 0.038 & $-$60.5 $\pm$ 2.6 & $-$1.13$\pm$0.2  & 2.84 \\
4050671278572353536 & 273.508670 & $-$28.614204 & 3.254 $\pm$ 0.042 & $-$3.387 $\pm$ 0.033 &  $-$65.3 $\pm$ 1.1 & $-$1.11$\pm$0.2  & 1.31 \\
4050624205664391040$^{\rm HB}$ &  273.537938 & $-$28.638492 & 2.811 $\pm$ 0.145 & $-$3.983 $\pm$ 0.109 & $-$48.8  $\pm$ 6.7 & - & 1.38 \\
4050624274430501248$^{\rm HB}$ &  273.527754 & $-$28.640416 & 3.695 $\pm$ 0.057 & $-$3.623 $\pm$ 0.043 & $-$54.1 $\pm$ 5.6 & - & 0.83 \\
4050647707732147456$^{\rm HB}$ &  273.493876 & $-$28.651595 &  3.878 $\pm$ 0.093 & $-$3.358 $\pm$ 0.073 & $-$46.8 $\pm$ 9.5 & - & 1.61 \\
4050647772223823488 & 273.484421 & $-$28.625506 & 3.982 $\pm$ 0.085 & $-$3.284 $\pm$ 0.071 & $-$74.3 $\pm$ 1.3 & $-$1.37$\pm$0.2  & 1.93 \\
4050600806719928576 & 273.517931 & $-$28.646514 & 3.483 $\pm$ 0.059 & $-$3.951 $\pm$ 0.048 & $-$62.3 $\pm$ 6.6 & - & 0.71 \\
4050624308743727744 & 273.536860 & $-$28.623270 & 3.364 $\pm$ 0.064 & $-$3.453 $\pm$ 0.050 & $-$46.2 $\pm$ 9.9 & - & 1.49 \\
4050647669052101248 & 273.473132 & $-$28.636323 & 3.531 $\pm$ 0.029 & $-$3.626 $\pm$ 0.024 & $-$49.1 $\pm$ 4.5 & $-$1.07$\pm$0.2  & 2.52 \\
4050671823952489088 & 273.549330 & $-$28.573736 & 3.558 $\pm$ 0.035 & $-$3.585 $\pm$ 0.027 & $-$59.5 $\pm$ 3.2 & - & 4.22 \\
4050624270079129216$^{\rm RR}$ & 273.520749 & $-$28.633437 & 3.762 $\pm$ 0.052 & $-$3.761 $\pm$ 0.040 & $-$39.8$\pm$12.4 & - & 0.36 \\
\hline
$^{\rm HB}$ Horizontal Branch Star & & & & & & & \\
$^{\rm RR}$ RR Lyrae Star & & & & & & & \\
\enddata
\end{deluxetable*}

\begin{deluxetable*}{lccccccc}
\tabletypesize{\scriptsize}
\tablenum{3}
\tablecaption{Positions, $Gaia$ proper motions, radial velocities and spectroscopic 
$\rm [Fe/H]$ metallicities of the candidate extra-tidal stars stripped from BH~261
\label{tab:ext_rvfe}}
\tablehead{
\colhead{{\it Gaia} ID} & \colhead{RA (deg)} & \colhead{Dec (deg)} & \colhead{$\mu_{\alpha}$ (mas~s$^{-1}$)} & \colhead{$\mu_{\delta}$ (mas~s$^{-1}$)} & \colhead{HRV (km~s$^{-1}$)} & \colhead{$\rm [Fe/H]$} & \colhead{r (') }
}
%\decimalcolnumbers
\startdata
Red Clump Stars &  &  &  &  & &  &  \\
\hline
4050678902240441728 & 273.47989 & -28.41042 & 3.117 $\pm$ 0.060 & -4.065 $\pm$ 0.040 & -76.4 $\pm$ 3.2 & -1.02 & 13.639 \\ 
4050583038419373568 & 273.66186 & -28.83802 & 3.754 $\pm$ 0.048 & -5.041 $\pm$ 0.035 & -66.5 $\pm$ 4.3 & -0.96 & 15.034 \\ 
4050609598466008576 & 273.86595 & -28.64076 & 2.418 $\pm$ 0.065 & -2.225 $\pm$ 0.048 & -40.3 $\pm$ 1.9 & -0.54 & 21.051 \\ 
4050637704736314496 & 273.17874 & -28.86039 & 2.854 $\pm$ 0.092 & -4.043 $\pm$ 0.065 & -39.2 $\pm$ 2.4 & -0.64 & 24.289 \\ 
4050611595682449152 & 274.02483 & -28.67061 & 3.054 $\pm$ 0.043 & -4.090 $\pm$ 0.032 & -41.6 $\pm$ 2.0 & -0.44 & 30.664 \\ 
4050859531382762880 & 273.02371 & -28.29882 & 4.076 $\pm$ 0.090 & -2.628 $\pm$ 0.068 & -50.8 $\pm$ 4.6 & -1.10 & 35.719 \\ 
4049822966614101376 & 273.83060 & -29.16159 & 3.827 $\pm$ 0.068 & -3.117 $\pm$ 0.049 & -50.2 $\pm$ 3.3 & -0.61 & 36.834 \\ 
4050863826171182336 & 272.95921 & -28.21074 & 3.688 $\pm$ 0.051 & -2.835 $\pm$ 0.037 & -72.2 $\pm$ 2.0 & -0.66 & 41.954 \\ 
4052192933885824896 & 274.04220 & -28.11325 & 3.578 $\pm$ 0.048 & -3.272 $\pm$ 0.039 & -56.2 $\pm$ 1.7 & -0.61 & 44.505 \\ 
4050770986475829120 & 272.56501 & -28.47638 & 4.074 $\pm$ 0.097 & -4.250 $\pm$ 0.069 & -55.1 $\pm$ 1.8 & -0.99 & 57.790 \\ 
4049707659748737920 & 273.46234 & -29.62002 & 3.072 $\pm$ 0.059 & -4.050 $\pm$ 0.041 & -59.4 $\pm$ 1.8 & -0.83 & 59.186 \\ 
4052214473131206400 & 273.95238 & -27.74946 & 3.838 $\pm$ 0.041 & -3.343 $\pm$ 0.030 & -71.1 $\pm$ 1.9 & -0.68 & 59.260 \\ 
4052143142313784960 & 274.37092 & -28.11093 & 3.090 $\pm$ 0.043 & -3.096 $\pm$ 0.031 & -54.7 $\pm$ 1.7 & -0.86 & 60.216 \\ 
\hline
Giant Stars &  &  &  &  & &  &  \\
\hline
4050622105410855296 & 273.70971 & -28.61684 & 2.749 $\pm$ 0.039 & -3.637 $\pm$ 0.030 & -77.0 $\pm$ 1.8 & -0.98 & 11.731 \\ 
4050619425373339008 & 273.74356 & -28.69270 & 2.780 $\pm$ 0.029 & -3.489 $\pm$ 0.020 & -36.2 $\pm$ 1.7 & -1.51 & 14.143 \\ 
4050681857034868608 & 273.70796 & -28.33282 & 3.180 $\pm$ 0.038 & -5.003 $\pm$ 0.028 & -54.4 $\pm$ 1.4 & -- & 21.511 \\
4050642828694691584 & 273.17722 & -28.76563 & 2.515 $\pm$ 0.042 & -3.843 $\pm$ 0.032 & -44.2 $\pm$ 1.7 & -0.55 & 21.728 \\ 
4050698113539687680 & 273.51378 & -28.22135 & 3.185 $\pm$ 0.085 & -3.431 $\pm$ 0.063 & -72.1 $\pm$ 2.2 & -0.80 & 24.820 \\ 
4050695703975916416 & 273.33456 & -28.24572 & 4.126 $\pm$ 0.040 & -2.934 $\pm$ 0.028 & -36.6 $\pm$ 1.9 & -0.62 & 25.746 \\ 
\hline
\enddata
\end{deluxetable*}

\clearpage

\begin{acknowledgments}
AMK acknowledges support from grant 
AST-2009836 from the National Science Foundation.  The grant support provided, in part, by the 
M.J. Murdock Charitable Trust (NS-2017321) is acknowledged.  
This work was made possible through the Preparing for
Astrophysics with LSST Program, supported by the Heising-Simons Foundation and
managed by Las Cumbres Observatory.  
M.J. gratefully acknowledges funding of MATISSE: \textit{Measuring Ages Through Isochrones, Seismology, and Stellar Evolution}, awarded through the European 
Commission's Widening Fellowship. This project has received funding from the European Union's Horizon 2020 research and innovation programme.

This project used data obtained with the Dark Energy Camera (DECam), which was constructed by the Dark Energy Survey (DES) collaboration. Funding for the DES Projects has been provided by the US Department of Energy, the US National Science Foundation, the Ministry of Science and Education of Spain, the Science and Technology Facilities Council of the United Kingdom, the Higher Education Funding Council for England, the National Center for Supercomputing Applications at the University of Illinois at Urbana-Champaign, the Kavli Institute for Cosmological Physics at the University of Chicago, Center for Cosmology and Astro-Particle Physics at the Ohio State University, the Mitchell Institute for Fundamental Physics and Astronomy at Texas A\&M University, Financiadora de Estudos e Projetos, Fundação Carlos Chagas Filho de Amparo à Pesquisa do Estado do Rio de Janeiro, Conselho Nacional de Desenvolvimento Científico e Tecnológico and the Ministério da Ciência, Tecnologia e Inovação, the Deutsche Forschungsgemeinschaft and the Collaborating Institutions in the Dark Energy Survey.

The Collaborating Institutions are Argonne National Laboratory, the University of California at Santa Cruz, the University of Cambridge, Centro de Investigaciones En\'{e}rgeticas, Medioambientales y Tecnol\'{o}gicas–Madrid, the University of Chicago, University College London, the DES-Brazil Consortium, the University of Edinburgh, the Eidgen\"{o}ssische Technische Hochschule (ETH) Z\"{u}rich, Fermi National Accelerator Laboratory, the University of Illinois at Urbana-Champaign, the Institut de Ci\'{e}ncies de l’Espai (IEEC/CSIC), the Institut de Física d’Altes Energies, Lawrence Berkeley National Laboratory, the Ludwig-Maximilians Universität München and the associated Excellence Cluster Universe, the University of Michigan, NSF’s NOIRLab, the University of Nottingham, the Ohio State University, the OzDES Membership Consortium, the University of Pennsylvania, the University of Portsmouth, SLAC National Accelerator Laboratory, Stanford University, the University of Sussex, and Texas A\&M University.

Based on observations at Cerro Tololo Inter-American Observatory, NSF’s NOIRLab (NOIRLab Prop. ID 2013A-0529; 2014A-0480; PI: M. Rich), which is managed by the Association of Universities for Research in Astronomy (AURA) under a cooperative agreement with the National Science Foundation.

This work has made use of data from the European Space Agency (ESA) mission {\it Gaia} (\url{https://www.cosmos.esa.int/gaia}), processed by the {\it Gaia} Data Processing and Analysis Consortium (DPAC, \url{https://www.cosmos.esa.int/web/gaia/dpac/consortium}). Funding for the DPAC has been provided by national institutions, in particular the institutions participating in the {\it Gaia} Multilateral Agreement.

\end{acknowledgments}

{}

\end{document}